\documentclass[iop]{emulateapj}

\usepackage{amsmath}
\usepackage{natbib}
\usepackage{graphicx}
\shorttitle{Q0302}
\shortauthors{Syphers \& Shull}

\begin{document}

\title{{\it HST}/COS Observations of the Quasar Q0302$-$003: Probing the \ion{He}{2} Reionization Epoch and QSO Proximity Effects}

\author{David Syphers and
J.\ Michael Shull
}
\affil{CASA, Department of Astrophysical and Planetary Sciences, University of Colorado, Boulder, CO 80309, USA}

\email{David.Syphers@colorado.edu}  

\begin{abstract}
Q0302$-$003 ($z=3.2860 \pm 0.0005$) was the first quasar discovered that showed a \ion{He}{2} Gunn-Peterson trough, a sign of incomplete helium reionization at $z \gtrsim 2.9$.
We present its {\it HST}/Cosmic Origins Spectrograph far-UV medium-resolution spectrum, which resolves many spectral features for the first time, allowing study of the quasar itself, the intergalactic medium, and quasar proximity effects.
Q0302$-$003 has a harder intrinsic extreme-UV spectral index than previously claimed, as determined from both a direct fit to the spectrum (yielding $\alpha_{\nu} \approx -0.8$) and the helium-to-hydrogen ion ratio in the quasar's line-of-sight proximity zone.
Intergalactic absorption along this sightline shows that the helium Gunn-Peterson trough is largely black in the range $2.87 < z < 3.20$, apart from ionization due to local sources, indicating that helium reionization has not completed at these redshifts.
However, we tentatively report a detection of nonzero flux in the high-redshift trough when looking at low-density regions, but zero flux in higher-density regions.
This constrains the \ion{He}{2} fraction to be about 1\% in the low-density IGM and possibly a factor of a few higher in the IGM as a whole, suggesting helium reionization has progressed substantially by $z \sim 3.1$.
The Gunn-Peterson trough recovers to a \ion{He}{2} Ly$\alpha$ forest at $z < 2.87$.
We confirm a transmission feature due to the ionization zone around a $z=3.05$ quasar just off the sightline, and resolve the feature for the first time.
We discover a similar such feature possibly caused by a luminous $z=3.23$ quasar further from the sightline, which suggests that this quasar has been luminous for $>$34~Myr.

\end{abstract}

\keywords{galaxies: active --- intergalactic medium --- quasars: absorption lines --- quasars: individual (Q0302$-$003) --- ultraviolet: galaxies}

\section{Introduction}
\label{sec:intro}
The full reionization of helium at $z \sim 3$ was the last major phase change of the intergalactic medium (IGM).
Because stars do not produce significant quantities of photons with high enough energies to ionize \ion{He}{2} ($\lambda < 228$~\AA), this reionization occurs only when there are enough photons from quasars.
Widespread helium reionization began at redshifts $z \gtrsim 3.5$, perhaps at $z \gtrsim 4$ \citep{becker11}.

The earlier stages of helium reionization can be probed through the \ion{H}{1} Ly$\alpha$ forest, observing the IGM heating that occurs during reionization \citep{becker11,bolton12}.
Observing metals can provide more certain constraints by breaking the degeneracy between thermal and nonthermal line broadening, but simulations indicate the optimal density regime for this can only be probed by future $\sim$30~m telescopes \citep{meiksin10}.
Earlier hopes that the evolution of the optical depth of the \ion{H}{1} Ly$\alpha$ forest would contain a clear imprint from helium reionization \citep[e.g.,][]{theuns02,bernardi03} have been overturned theoretically and observationally \citep{bolton09a,becker13}.
A damping red wing of IGM absorption can be present at high \ion{He}{2} fractions, and while quite challenging to detect, it may be possible \citep{syphers11a}.

The most direct way to study the later stages of helium reionization is via the \ion{He}{2} Gunn-Peterson trough, where \ion{He}{2}~Ly$\alpha$ absorption (303.7822~\AA) at every redshift in some range leads to a black trough.
This Gunn-Peterson trough is difficult to see, because intervening hydrogen is optically thick for $\sim$95\% of $z \sim 3$ quasars \citep{picard93,zheng05}.
Those few quasars in which the trough can be seen are called \ion{He}{2} quasars.
In recent years, many new \ion{He}{2} quasars have been discovered \citep{syphers09a,syphers09b,syphers12,worseck11a}.
The most basic statistic that can be examined is the effective optical depth and its evolution with redshift, $\tau_{\rm eff}(z)$ \citep[e.g.,][]{syphers11a}.
Some theoretical work indicated this could fruitfully probe well into the helium reionization epoch \citep{dixon09}, but more recently it has been realized that $\tau_{\rm eff}(z)$ can depend strongly on factors unrelated to reionization \citep{khaire13,davies14}.
As a result, arguments based on fluctuations, dark gaps, and more sophisticated measurements are now being pursued.

Our current work examines Q0302$-$003 (henceforth Q0302).
This was the first \ion{He}{2} quasar discovered, its \ion{He}{2} break seen in an {\it HST}/Faint Object Camera (FOC) spectrum \citep{jakobsen94}.
This sightline has also been studied at low resolution ($R \equiv \lambda / \Delta \lambda \sim 1000$) with Goddard High-Resolution Spectrograph and Space Telescope Imaging Spectrograph \citep[GHRS and STIS;][]{hogan97,heap00}.
Q0302 is not the brightest \ion{He}{2} quasar, but those much brighter are all at lower redshift, at the tail end of helium reionization.
Most prominently, these include HS1700$+$6416 at $z=2.75$ \citep{davidsen96,syphers13}, 4C57.27 at $z=2.86$ \citep{syphers12}, and HE2347$-$4342 at $z=2.89$ \citep{reimers97,shull10}.
The brightness and relatively high redshift of Q0302 is surpassed only by SDSS0915 \citep{syphers12}, and our team is currently analyzing new observations of the latter to compare with our Q0302 results.

We detail the observations in Section~\ref{sec:obs}.
In Section~\ref{sec:q0302_properties}, we examine Q0302 itself, fitting its continuum, determining its systemic redshift, and discussing possible intrinsic quasar emission.
In Section~\ref{sec:tau}, we consider the \ion{He}{2} optical depth, its redshift evolution and its possible dependence on density (as inferred from the \ion{H}{1}~Ly$\alpha$ forest).
In Section~\ref{sec:eta}, we use comparisons of \ion{He}{2} and \ion{H}{1} absorption to examine the IGM-ionizing background and its fluctuations during both the pre and post-reionization eras.
In Section~\ref{sec:proximity} we discuss line-of-sight and transverse proximity effects seen in the Q0302 sightline.
Such proximity effects arise in regions near quasars, where the transmitted flux is affected by ionizing radiation that is increased or harder compared to the metagalactic background \citep[e.g.,][]{bajtlik88,crotts89}.
We conclude in Section~\ref{sec:conclusion}.
In Appendix~\ref{app:eta}, we discuss methodological details of measuring fluctuations in the ionizing background.
In Appendix~\ref{app:continuum}, we comment on the problem of continuum placement in the normalization of \ion{H}{1} data.
Appendix~\ref{app:tab_data} presents tabulated data for the optical depth figure.

Throughout the paper we use a WMAP9 cosmology \citep{hinshaw13}, and a primordial helium mass fraction $Y_p=0.2485$ \citep[][with WMAP9 cosmological parameters]{steigman07}, which is in agreement with recent $Y_p$ determinations by independent methods \citep{aver13}.

\section{Observations}
\label{sec:obs}

The new FUV observations at the core of this paper are from the Cosmic Origins Spectrograph aboard {\it HST} \citep[COS;][]{green12}.
We also present a new NIR spectrum of Q0302 from the Apache Point Observatory (APO) 3.5~m, to determine the systemic quasar redshift.
In addition, several archival data sets are used.
All the spectroscopic observations of Q0302 used in this paper are detailed in Table~\ref{tab:obs}, and described in more detail below.

\subsection{{\it HST} FUV Data}
\label{sec:fuv_data}

\begin{figure}
\epsscale{1.2}
\plotone{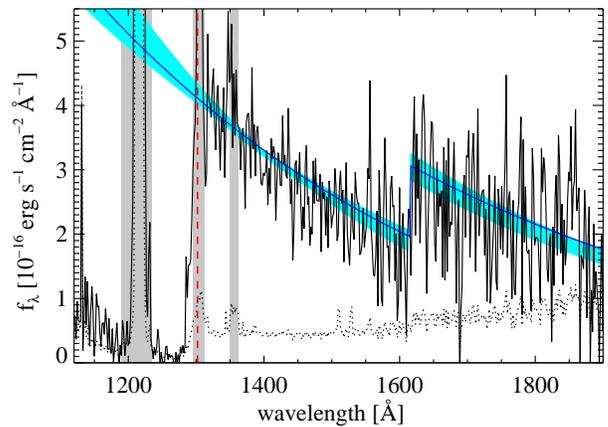}
\caption{Dereddened COS G140L spectrum (solid black line) with uncertainty (dotted black line) and the continuum fit overplotted (blue line), and the continuum uncertainty shown (cyan shading). The spectrum is binned to 1.7~\AA, three resolution elements. The flux break from IGM \ion{He}{2}~Ly$\alpha$ is shown as the vertical dashed red line. The continuum fit includes all partial Lyman-limit systems (pLLS) in Table~\ref{tab:lls}, although only COS/G140L data are used to determine the spectral index and normalization. Geocoronal contamination is evident for Ly$\alpha$, \ion{O}{1}~$\lambda$1304, and possibly \ion{O}{1}]~$\lambda$1356 (gray shaded regions).}
\label{fig:cos_full_continuum}
\end{figure}

\begin{figure}
\epsscale{1.2}
\plotone{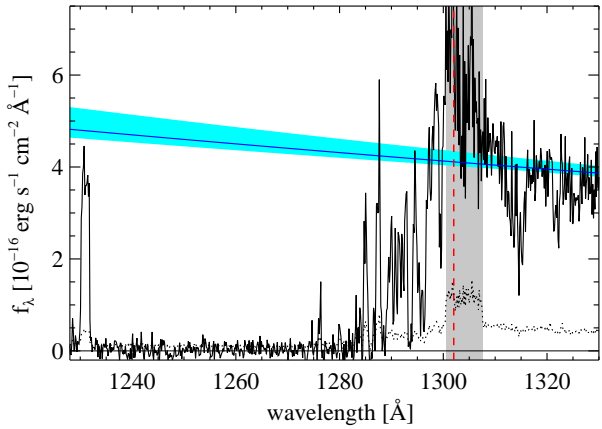}
\caption{COS G130M data near the \ion{He}{2}~Ly$\alpha$ edge (vertical dashed red line). The spectrum fit is overplotted in blue, with the continuum uncertainty shaded in cyan. The dotted line shows spectrum uncertainty. The spectrum is binned to 0.14~\AA, two resolution elements. Note that the continuum fit, while derived only using the COS G140L data, matches the G130M data very well at $\lambda > 1315$~\AA. Blueward of this, geocoronal \ion{O}{1} emission contaminates the spectrum (gray shaded region), despite using night-only data here. In addition, some real quasar line emission may be evident, as discussed in Section~\ref{sec:intrinsic_emission}.}
\label{fig:cos_zoomin}
\end{figure}

Q0302 was observed for 24.87~ks with COS/G130M and 2.67~ks with COS/G140L, in GO~12033.
The G140L spectrum is shown in Figure~\ref{fig:cos_full_continuum}, and the portion of the G130M spectrum covering the \ion{He}{2} break is shown in Figure~\ref{fig:cos_zoomin}.
In order to achieve continuous spectral coverage across the  G130M bandpass (1135~\AA\  $\lesssim\lambda \lesssim$ 1460~\AA) and minimize fixed-pattern noise, we made observations at two central wavelength settings (1291~\AA\ and 1300~\AA) with four focal-plane offset locations in each grating setting.
The data were reduced using CALCOS 2.18.5, although with some custom modifications described below.
The COS data were taken at the original lifetime position of COS, but only a few months before its shift to the second lifetime position.
Thus compared to more recent observations we have slightly higher resolution (7 pixels FWHM, or $R \approx 18$,000 at 1250~\AA, instead of the 8 pixels after the first lifetime position adjustment) but are affected nontrivially by loss of detector sensitivity, discussed in detail in the appendix of \citet{syphers12}.

Geocoronal emission lines contaminate FUV spectra at certain wavelengths, although any scattered light background is negligible.
In an attempt to reduce geocoronal emission, we use only data taken during orbital night in those spectral regions affected.
There are 6.35~ks of night data for Q0302, but unusually strong geocoronal emission from \ion{O}{1}~$\lambda$1304 and \ion{O}{1}]~$\lambda$1356 is visible for this target, including for when the Sun is at an angle that usually leaves no observable contamination.
We cut the data more aggressively than usual to avoid this contamination, but faint residual contamination is still visible, and we point it out when it impacts our analysis.
Unless otherwise specified, all COS spectra in this paper use night-only data in wavelength regions affected by geocoronal contamination, and all data in other  regions.

\begin{deluxetable*}{llllll}
\tablewidth{0pc}
\tabletypesize{\footnotesize}
\tablecaption{Spectroscopic Observations of Q0302$-$003}
\tablehead{
\colhead{Observatory} & \colhead{Instrument} & \colhead{Exp.~Time} & \colhead{Wavelength Range} & \colhead{Resolution} & \colhead{Date} \\
\colhead{} & \colhead{} & \colhead{(ks)} & \colhead{(\AA)} & \colhead{} & \colhead{}}
\startdata
HST & COS/G140L & \phn2.67 & 1120--1900 & 2500\tablenotemark{a} & 2012 Mar 9\\
HST & COS/G130M & 24.87 & 1135--1450 & 18,000\tablenotemark{a} & 2012 Mar 8--9\\
HST & STIS/G140L & 23.28 & 1150--1720 & 750 & 1997 Dec 2, 10\\
HST & STIS/G230L & \phn3.69 & 1600--3150 & 500 & 1999 Oct 2 \\
APO & TripleSpec & 13.80 & 0.95--2.46~$\mu$m & 3500 & 2013 Jan 19 \\
Keck & HIRES & 27.00 & 4220--6660 & 43,000 & 1998 Jan 29--30 \\
VLT & UVES & 39.90 & 4810--5770 & 45,000 & 1999 Oct 12, 15, 16\\
\enddata
\tablenotetext{a}{The COS line-spread function (LSF) has substantial non-Gaussian wings. Throughout this work we call seven pixels a COS resolution element, which accurately reflects the real LSF FWHM, but does not account for the extent of these wings.}
\label{tab:obs}
\end{deluxetable*}

The background of COS is better understood than that of earlier FUV detectors, and lower than many, including {\it FUSE}.
Nonetheless, for faint targets and for highly absorbed regions, such as Gunn-Peterson observations of Q0302, the background of COS is a substantial issue.
The background is particularly large for the higher-resolution observations we use in the present work, since it is constant per pixel.
(STIS/G140L observations have lower per-{\AA}ngstrom dark current than COS/G130M, due to the lower resolution, and for early STIS observations such as those used here, it was much lower. However, in addition to the benefits of higher resolution, COS also has much higher sensitivity.)
We improve on the pipeline background subtraction in three ways.

First, we choose stringent pulse-height amplitude (PHA) limits on counts we accept, because source counts and background counts have different distributions.
The CALCOS default changed in December 2012 from including PHA$=$2--30 to including PHA$=$2--23.
However, this default is still conservative, as it is intended to work with all data sets, and we find for the Q0302 observations that PHA exclusion can be more aggressive.
For G130M segment A, we use PHA$=$1--9, which includes $>$99\% of all source counts.
A few source counts remain at PHA$=$0, but none are evident at PHA$>$9.
G130M segment B has relatively few counts due to the Gunn--Peterson trough, but we can still set good PHA limits, using PHA$=$2--15.
This agrees with previous work indicating that the PHAs are shifted to higher values by the higher voltage of this segment, but there are also low-PHA regions due to high gain sag from the geocoronal Ly$\alpha$ line burn-in \citep{syphers13}.
For G140L, only segment A is used, and we prefer PHA$=$1--12.

Second, we use a modified primary science aperture (PSA).
The default PSA is 35 pixels in the cross-dispersion direction for G130M and 57 pixels for G140L.
However, this is extremely conservative, a large overestimate of the spectrum width.
For G130M segment A we rigorously derive a PSA width of 20 pixels is all that is needed to include essentially all flux.
Segment B has fewer counts, and so cannot be constrained as precisely, but checks there indicate 20 pixels is fine for this segment as well.
G140L suggests 15 pixels would include essentially all source flux, but we are unable to verify this near the edges of the detector due to low fluxes and sensitivities, and we know the spectrum spreads some at the ends \citep{syphers13}.
As a result, we use a PSA cross-dispersion width of 20 pixels for all observation modes.

Third, we use a background estimate from the PSA itself.
The pipeline estimate of the background uses from regions offset in the cross-dispersion direction, which are not exposed to source photons.
However, the background is not constant across the entire detector, and one particularly large effect here is burn-in.
The COS detector loses sensitivity over time when exposed to light, so the PSA has reduced sensitivity compared to regions that are not illuminated.
The improved background subtraction method we suggested and then implemented in \citet{syphers12} and \citet{syphers13} is required for Q0302 data.
In this method we use data from COS dark monitoring campaigns, which directly measure the background in the PSA.
By adding many dark exposures taken close to the same time the science data were taken, we obtain an estimate of the true dark rate in the PSA.

Wavelength calibration can be a serious issue for some COS modes \citep{syphers13}, but the modes used in this observation are well calibrated.
We have verified the calibration of the G130M data on segment A to within one resolution element using Galactic interstellar \ion{C}{2}~$\lambda$1334.53~\AA\ and \ion{C}{2}*~$\lambda$1335.71~\AA, which given our S/N and wavelength coverage are the only Galactic lines easily measured.
Segment B has no usable Galactic lines, but in the line-of-sight proximity zone of the quasar we can verify that the \ion{H}{1} and \ion{He}{2} Ly$\alpha$ absorption features align to within a single COS resolution element ($\lesssim$17~km~s$^{-1}$).

\subsection{Comparisons to Earlier FUV Data and Results}
\label{sec:comparisons}
Q0302 was observed with STIS G230L for 3.7~ks (GO 7272, PI Hogan) and G140L 21.6~ks (GO 7575, PI Heap).
We use G140L from 1150--1720~\AA, and G230L data from 1720--3150~\AA.
For a more detailed discussion of the STIS data, we refer to \citet{heap00}.
The STIS spectrum presented in this paper is the default CALSTIS reduction of the data (version 2.22 for FUV, 2.23 for NUV).
We note that the default background subtraction is untrustworthy.
Improved background subtraction is possible \citep{heap00}, but as we are interested only in the unabsorbed continuum and differential features in the Gunn-Peterson trough, we do not make the effort to improve the smooth background determination.

\begin{figure}
\epsscale{1.2}
\plotone{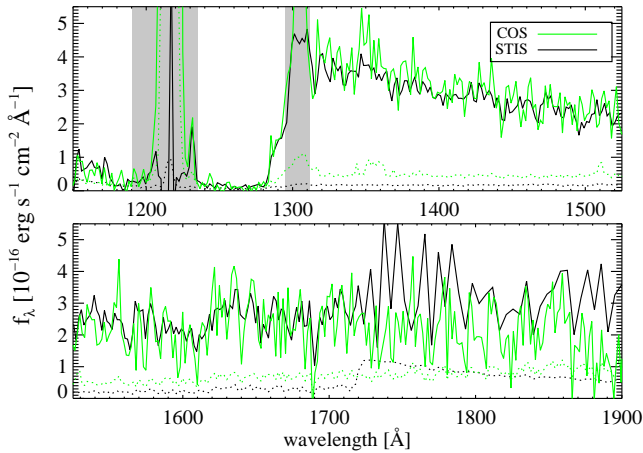}
\caption{COS G140L (green) overplotted on the STIS data (black; G140L at $\lambda < 1720$~\AA\ and G230L at $\lambda > 1720$~\AA). The COS spectrum is binned to three resolution elements (1.7~\AA) and the STIS spectra are binned to about one resolution element (1.8~\AA\ G140L, 4.6~\AA\ G230L). The flux levels observed in the COS data are consistent with those in the STIS G140L data (taken $3.3$ years earlier in the quasar frame). The trend of the STIS G230L data does not match; while this could be quasar variability, it could also be an instrumental artifact of the STIS data. Regions with some geocoronal contamination in the COS data are shaded in gray.}
\label{fig:cos_stis}
\end{figure}

The nominal resolution of the data is $R \approx $~1200--2000 for STIS G140L, but the use of the somewhat wider $0 \farcs 2$ slit does increase power in the wings (although it leaves the FWHM essentially unchanged).
As a result, \citet{heap00} claim a spectral resolution of only 3~pixels, or $R \approx $~700--1000.
A comparison to COS FUV is shown in Figure~\ref{fig:cos_stis}.
The slight mismatch in fluxes in the Gunn-Peterson trough is spurious, and due to the background issue mentioned above.
The COS and STIS FUV continuum flux levels match very well despite independent normalization.

We revisit several literature results based on STIS data in subsequent sections, but here we consider one example, a detection of a \ion{He}{1}~$\lambda$584 absorption line.
\ion{He}{1}~$\lambda$584 is seen extremely rarely in the IGM; we are aware of detection claims in the literature only in the sightlines of HS1700$+$6416 \citep{reimers93,syphers13} and Q0302.
\citet{heap00} claim a likely $z=1.890$ \ion{He}{1}~$\lambda$584 detection at 1689~\AA\ in the Q0302 STIS spectrum.
We confirm the reality of this absorption feature, and measure an observed-frame EW~$\sim$~2.5~\AA.
\citet{hu95} observe a \ion{C}{4} and \ion{Al}{2} system at $z=1.8924$.
Although we do not have coverage of the \ion{H}{1} Lyman-series lines in our data, we do not observe a Lyman limit break associated with this; see Section~\ref{sec:specfit}.

The two \ion{He}{1}~$\lambda$584 lines in HS1700$+$6416 are associated with absorption systems of column density $\log{N_{\rm H \, I}} \approx 16.0$ and $16.6$ \citep{fechner06a}.
We find the identification of the feature in the Q0302 spectrum as \ion{He}{1}~$\lambda$584 plausible, and absent detailed modeling of this absorption system, more likely than other, even more rare lines (such as \ion{Ne}{6}~$\lambda$559 associated with the $z=2.015$ system).
The absorption appears to be 1~\AA\ ($\sim$200~km~s$^{-1}$) off from what one would expect if it were associated with the metal-line system, but this alignment is still better than other possible line identifications.
This difference is less than twice the velocity resolution of COS/G140L, and so it should not be regarded as highly significant.

Another obvious possibility is that the 1689~\AA\ absorption line is a low-redshift hydrogen Lyman-series line, but we regard this as unlikely.
We can rule out Ly$\gamma$ and higher from the lack of other stronger Lyman-series lines in the COS spectrum at longer wavelengths, as well as a lack of a Lyman break.
We can rule out Ly$\alpha$ and Ly$\beta$ because the implied rest-frame equivalent widths ($W_{\alpha} \sim 1.8$~\AA, $W_{\beta} \sim 1.5$~\AA) have $N_{\rm H \, I}$ large enough to cause an observable Lyman break.
Even if the break occurred in the Gunn-Peterson trough, the flux would be depressed enough that we would not see the recovery to a \ion{He}{2}~Ly$\alpha$ forest.
In addition, if the line were Ly$\alpha$, it would cause easily observable Ly$\beta$ in our G130M spectrum, which we do not see.
This would be the case even if the absorption is actually a blend of a small number of forest systems.

\subsection{Ground-Based Observations}

Many unique opportunities are possible with \ion{He}{2} quasars, because one can have simultaneous information on both the hydrogen and the helium IGM at every redshift.
To make use of that we need high-resolution optical data on the \ion{H}{1}~Ly$\alpha$ forest at the redshifts of interest.
For this we use the VLT/UVES spectrum \citep{kim02} and a Keck/HIRES spectrum, kindly provided to us in normalized form by Tae-Sun Kim and George Becker, respectively.
The UVES spectrum has higher S/N, and we use that where possible, although it does not cover the entire region between Ly$\alpha$ and Ly$\beta$.

For our analysis, knowing the systemic redshift of Q0302 is important.
As we discuss in Section~\ref{sec:redshift}, the best way to do this is with emission lines that occur in the observed-frame near IR (NIR).
We took a NIR spectrum with the 3.5~m at the APO, using TripleSpec, which has coverage across J, H, and K bands at a resolution of $R \sim 3500$.
The resulting spectrum consists of 230 minutes of integration on target, taken with a $1 \farcs 1$ slit at temperatures $\sim$2~$^{\circ}$C.
Light clouds were occasionally present, but they do not affect our scientific goal of line centroiding.
We reduced the data using TripleSpecTool, a modified version of SpexTool \citep{cushing04}.
Because we are using this to determine a precise redshift, the wavelength calibration is important.
We calibrated the wavelength using both atmospheric OH lines and a correction based on a NeAr lamp spectrum, and corrected the spectrum to heliocentric velocities, achieving a velocity error of $\lesssim$15~km~s$^{-1}$ in all bands (better than 10~km~s$^{-1}$ in K~band).

\section{Properties of Q0302}
\label{sec:q0302_properties}

\subsection{Fitting the Continuum}
\label{sec:specfit}

Prior to fitting the COS spectrum, we correct for Galactic extinction using $E(B-V)=0.081$ \citep{schlafly11}, $R_V=3.1$, and the extinction curve of \citet{fitzpatrick99}.
This extinction is significant for observations in the FUV, and is the highest of any known \ion{He}{2} quasar.
However, uncertainties in $E(B-V)$ do not affect our primary science goal of optical depth analysis, as the reddening is divided out of flux ratios for nearby wavelengths.
Although choosing an average $R_V$ and this specific extinction curve leads to uncertainty and possibly some small bias \citep{peek13}, these errors are small compared to those discussed below even for fitting the spectral index.

We fit the continuum using the COS G140L data (Figure~\ref{fig:cos_full_continuum}), which, despite the lower resolution than G130M, are more useful here because they have much larger wavelength coverage.
We select regions of the continuum relatively free from absorption for an initial calculation of the continuum, and then iteratively select absorption-free windows by sigma clipping deviations of the spectrum from the fits, until we reach convergence.
We clip somewhat more aggressively on downward fluctuations than upward fluctuations, as the former include real absorption lines while the latter has only noise.
Choices such as binning and sigma clipping parameters dominate the uncertainty of the fit, and therefore we follow \citet{syphers12} by considering the full range of continua fit for reasonable parameter ranges as our continuum uncertainty (cyan shading in Figure~\ref{fig:cos_full_continuum}).
There are no quasar emission lines in this region large enough to affect our fit \citep{syphers12}.

It is standard to extrapolate the continuum as a pure power law for \ion{He}{2} quasars, but in those few cases where NUV data are available, we should fit the spectrum including known hydrogen absorption systems along the line of sight.
Although this may not always improve the extrapolation near the \ion{He}{2}~Ly$\alpha$ break \citep{syphers13}, it is more important for higher-redshift targets like Q0302, because the continuum must be extrapolated further.
In addition, it allows us to determine the intrinsic extreme-UV (EUV) spectral index of the quasar.

Many of the most common ion species observed in the IGM are created by photoionization from quasars, which also helps set the IGM temperature.
Nearly all of these ionization potentials, from \ion{H}{1} to \ion{C}{4}, are in the EUV, making knowledge of the quasar EUV spectral index important for modeling the IGM.
Since the EUV covers the ionization potential of \ion{He}{2}, knowing the properties of quasars here is also important for simulations of helium reionization.
Correction for host-galaxy extinction is unnecessary, because we are interested in what the IGM itself experiences.

The best-fit intrinsic FUV spectral index is $\alpha_{\nu}=-0.82$ (for $f_{\nu} \propto \nu^{\alpha_{\nu}}$), but the full range of spectral indices fit for a variety of smoothing lengths, sigma-clipping levels, and partial Lyman-limit system (pLLS) parameters is $\alpha_{\nu}=-0.37$ to $\alpha_{\nu}=-0.97$.
This is bluer than seen on average for luminous quasars \citep{telfer02,shull12}, but the best-fit value is within the normal range.
It is noticeably different from the $\alpha_{\nu}=-1.9$ fit to Q0302 by \citet{heap00}, but they do not correct for NUV pLLS (e.g., see their Figure~8).
They use optical and FUV data to constrain the slope, but without accounting for NUV pLLS, the depression of the FUV flux by IGM absorption will be seen as an intrinsically fainter FUV continuum.
This will yield a much softer slope than is actually the case, which likely accounts for the difference compared to our value.

There are five pLLS/LLS along this sightline (Table~\ref{tab:lls}).
The strongest one unfortunately occurs at a difficult wavelength, just above 3200~\AA, which is just redward of the STIS NUV coverage, and not covered in most optical spectra.
However, with very good blue optical coverage, \citet{sargent89} were able to observe this system, and found $\log{N_{\rm H \, I}}=17.38$.
This system shows \ion{C}{4} at $z=2.53531$ and $z=2.53563$, and a much weaker system at $z=2.53428$.
We adopt a redshift of $z=2.5355 \pm 0.0001$ for the LLS.

\begin{figure}
\epsscale{1.2}
\plotone{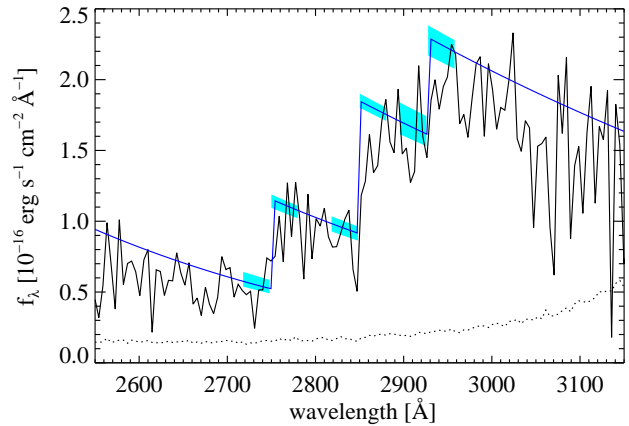}
\caption{STIS G230L data showing the three pLLS with Lyman limits covered, binned to one resolution element (4.6~\AA). The cyan shading shows the column density uncertainty reported in Table~\ref{tab:lls} for each individual system; this is not uncertainty in the overall spectral fit, the latter of which includes spectral index and normalization uncertainty. The blue line is the continuum as fit to the COS FUV data using the pLLS parameters from Table~\ref{tab:lls}, and scaled by a constant. See the text for a discussion of possible quasar variability, although this does not affect the derivation of the pLLS parameters.}
\label{fig:stis_plls}
\end{figure}

Three other systems have Lyman breaks covered by the STIS NUV data, at redshifts $z=2.015$, $2.125$, and $2.21$ (see Figure~\ref{fig:stis_plls}).
The extremely low resolution of STIS/G230L makes distinguishing the continuum above and below the Lyman break a poorly determined task, but we include uncertainty on the column densities in our continuum fit.
There is absorption consistent with both lines of \ion{C}{4} at $z=2.015$, no \ion{C}{4} seen at $z=2.125$, and strong \ion{C}{4} absorption at $2.22796 \pm 0.00007$.
For the last system, the STIS data prefer a break at $z \approx 2.21$; we use this redshift, and note that the precise redshift of the break makes no noticeable difference.
A fourth pLLS is observed in COS and STIS data at $z=0.771$, with $\log{N_{\rm H \, I}}=16.85^{+0.15}_{-0.10}$ measured in the COS data.
The much lower-resolution STIS data give a somewhat lower value, $\log{N_{\rm H \, I}}=16.65$.
We take the redshift from the Lyman break; unfortunately both the STIS and COS G140L data are too poor to distinguish any Lyman-series lines, and no metal lines appear in the optical data.
Some other \ion{C}{4} systems are observed, but none are associated with evident Lyman breaks.

Two other pLLS are possible from the STIS data, at $z \sim 0.68$ and $z \sim 0.86$.
However, the COS data rule out both, although the data are noisy enough to leave the possibility of a weak pLLS at the higher redshift.
The false appearance of a nontrivial pLLS ($\tau_{\rm LL} \sim 0.6$) at $z=0.86$ can be explained by noticing that the apparent break occurs at the transition between STIS FUV and NUV data, and with very high noise in NUV.
Although generally these two STIS modes match up well, this target is very faint for STIS, and backgrounds are considerable and not well subtracted \citep{heap00}.
In addition the two STIS data sets are not coeval, and thus quasar variability might play a role.
A $z \sim 0.60$ pLLS with relatively low $N_{\rm H \, I}$ is apparent in a continuum model of \citet{heap00}; it is unclear why, as it is not motivated by any evident drop in the STIS data.
As there is no hint of a drop in the COS data, we do not include it here.
(\citet{heap00} assumed $f_{\lambda}$ was {\it constant} for actual extrapolation to obtain optical depths.)

The possibility of quasar variability is an interesting one.
Figure~\ref{fig:stis_plls} uses a continuum fit with a power law and normalization constrained only by the COS FUV data and the pLLS parameters of Table~\ref{tab:lls}.
The spectral slope thus derived appears consistent with the STIS NUV data.
However, a constant normalization factor of $2.6$ was applied to the FUV-derived model spectrum before overplotting on the STIS NUV data.
This exact number is irrelevant to our analysis, but indeed the STIS G230L data overall appear 2--3 times higher than one would expect from the FUV flux.
The STIS G140L and G230L data sets were taken nearly two years apart (156 days apart in the quasar rest frame), so variability could explain the difference.
However, the COS and STIS G140L observations bracket the STIS G230L, and yet agree well with each other (Figure~\ref{fig:cos_stis}).

\begin{deluxetable}{lll}
\tablecolumns{9}
\tablewidth{0pc}
\tabletypesize{\footnotesize}
\tablecaption{LLS/pLLS in the Q0302$-$003 Sightline}
\tablehead{
\colhead{Redshift} & \colhead{$\log{N_{\rm H \, I}}$} & \colhead{$N_{\rm H \, I}$ Data} \\
\colhead{} & \colhead{(cm$^{-2}$)} & \colhead{}}
\startdata
$0.771 \pm 0.002$ & $16.85^{+0.15}_{-0.10}$ & COS G140L \\
$2.015 \pm 0.010$ & $17.10^{+0.05}_{-0.10}$ & STIS G230L \\
$2.125 \pm 0.005$ & $17.05^{+0.05}_{-0.05}$ & STIS G230L \\
$2.21 \phn \pm 0.02$\tablenotemark{a} & $16.75^{+0.10}_{-0.20}$ & STIS G230L \\
$2.5355 \pm 0.0001$ & 17.38 & Hale Double Spectrograph\tablenotemark{b} \\
\enddata
\tablenotetext{a}{The VLT spectrum shows a strong \ion{C}{4} absorber at $2.22796 \pm 0.00007$. The STIS data prefer a break at the lower redshift given here, but the exact redshift is unimportant for fitting the FUV continuum.}
\tablenotetext{b}{\citet{sargent89}; no error on $N_{\rm H \, I}$ is given.}
\label{tab:lls}
\end{deluxetable}

\subsubsection{\ion{He}{2}~Ly$\alpha$ Emission}
\label{sec:intrinsic_emission}

In addition to the intrinsic spectral index, any possible quasar line emission is of great interest.
Photoionization models predict strong \ion{He}{2}~Ly$\alpha$ emission \citep{syphers11a,lawrence12}, but it is rarely seen, and indeed has yet to be unequivocally established.
There is some broad emission likely seen in HS1700$+$6416 \citep{syphers13} and some low-resolution prism spectra \citep{syphers09a,syphers09b}, but uncertainties preclude definitive verification and identification until better data are obtained.
Unfortunately, the geocoronal emission in the Q0302 observations is unusually strong, and it cannot be entirely removed even when using night-only data.
The strongest \ion{O}{1} geocoronal line (1302.168~\AA) is nearly coincident in wavelength with the \ion{He}{2}~Ly$\alpha$ break ($1302.01 \pm 0.15$~\AA).
With two other strong geocoronal \ion{O}{1} lines at 1304.858~\AA\ and 1306.029~\AA, the contaminating influence covers most of $303.4$--$305.1$~\AA\ in the quasar rest frame, about 1200~km~s$^{-1}$ from the systemic center of \ion{He}{2}~Ly$\alpha$ on the red side.
The high resolution of COS G130M allows for the continuum to very briefly be recovered between the two bluer \ion{O}{1} lines, and there it is consistent with an extrapolation of the redder continuum.
There is, however, a small flux excess above the continuum seen just below the \ion{He}{2} break (Figure~\ref{fig:cos_zoomin}), which cannot be easily attributed to geocoronal contamination.
This could be weak evidence for intrinsic \ion{He}{2}~Ly$\alpha$ emission showing in the proximity zone, but without more of a line to fit, it remains highly tentative.

\subsection{Systemic Redshift of Q0302}
\label{sec:redshift}

The precise systemic redshift of Q0302 must be determined in order to properly analyze its proximity zone.
Uncertainty about the systemic redshift of a quasar can lead to dramatically different interpretations of the quasar near zone \citep[e.g.,][]{shull10}.
Since strong rest-frame FUV lines like \ion{C}{4}~1549 and \ion{H}{1}~Ly$\alpha$ are often substantially shifted from systemic \citep[e.g.,][]{vanden-berk01}, we pursue near-IR spectroscopy to access low-ionization and narrow forbidden lines.

\begin{figure}
\epsscale{1.2}
\plotone{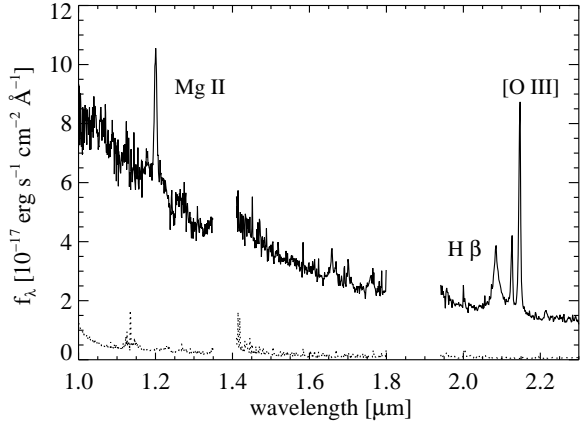}
\caption{Near-IR APO/TripleSpec spectrum of Q0302$-$003, prominently showing \ion{Mg}{2}~$\lambda$2799 in J~band, and H$\beta$ and [\ion{O}{3}]~$\lambda \lambda$4959/5007 in the K~band. We derive a systemic redshift of $z=3.2860 \pm 0.0005$. The dotted line is the spectrum error.}
\label{fig:NIR_spec}
\end{figure}

Using the lines shown in the APO/TripleSpec NIR spectrum (Figure~\ref{fig:NIR_spec}), we find a systemic $z=3.2860 \pm 0.0005$, which we adopt as our best estimate.
We fit \ion{Mg}{2}~$\lambda$2799, [\ion{Ne}{3}]~$\lambda\lambda$3869/3968, H$\beta$, and [\ion{O}{3}]~$\lambda\lambda$4959/5007, and use lab wavelengths for the lines.
If we use the \citet{vanden-berk01} rest wavelengths (corrected to [\ion{O}{3}]), this becomes $z=3.2870 \pm 0.0014$, due entirely to common substantial redshift of H$\beta$ and \ion{Mg}{2}, although more recent work suggests there is no offset $>$30~km~s$^{-1}$ for \ion{Mg}{2} \citep{hewett10}.
Although some of the lines are weak ([\ion{N}{3}]), and others are broad and hard to accurately centroid (H$\beta$), we can compare this value to the value from [\ion{O}{3}] $\lambda$5007 alone.
Because the line is asymmetric, we fit the top half of the line to find $z=3.2862$, in good agreement with the average.
The narrow component of H$\beta$, which is a good indicator of systemic velocity even when the broader component is shifted, gives $z=3.2857 \pm 0.0003$, also in agreement with the average.

We compare this redshift from the NIR lines to that obtained from the low-ionization line \ion{O}{1}~$\lambda$1304, observed in the optical.
Many \ion{He}{2} quasars have no NIR spectra, so it is of interest to see how accurate redshift determinations from optical spectra alone can be.
The \ion{O}{1}~$\lambda$1304 line in the SDSS spectrum is asymmetric, but when fitting its peak we find a redshift of $3.287 \pm 0.004$ when using the \citet{vanden-berk01} rest wavelength of 1305.42~\AA.
This becomes $3.293 \pm 0.002$, unacceptably large, if one uses the weighted laboratory average of 1303.49~\AA.
This agreement between \ion{O}{1}~$\lambda$1304 and the forbidden and low-ionization lines seen in the NIR matches what we see in other quasars examined.
HE2347$-$4342 has very good agreement between these (R.~Simcoe 2011, private communication), while HS1700$+$6416 has good but not perfect agreement \citep{syphers13}.
The weakness of the \ion{O}{1} line, the uncertainty over the proper rest wavelength, and the possibility of some offset do mean that where NIR spectra are practical, they are preferable for redshift determinations of \ion{He}{2} quasars.

\section{\ion{He}{2} Effective Optical Depth}
\label{sec:tau}

The simplest measurement possible of Gunn-Peterson data is the redshift evolution of effective optical depth.
There were originally hopes that this metric could constrain helium reionization \citep{dixon09,syphers11a}.
However, it now appears other effects could dominate over the reionization signal \citep{davies14}.

For comparison to other work, we report on the broad redshift evolution of the effective optical depth along the Q0302 sightline.
However, we also look at the more subtle and sensitive measure of contrasting $\tau_{\rm eff}$ in two different IGM density regimes.
Other methods include looking at dark gaps, where long dark troughs such as those seen in Q0302 are indicative of incomplete reionization \citep{furlanetto10}.

\subsection{Redshift Evolution of $\tau_{\rm eff}$}
\label{sec:tau_evolution}

\begin{figure}
\epsscale{1.2}
\plotone{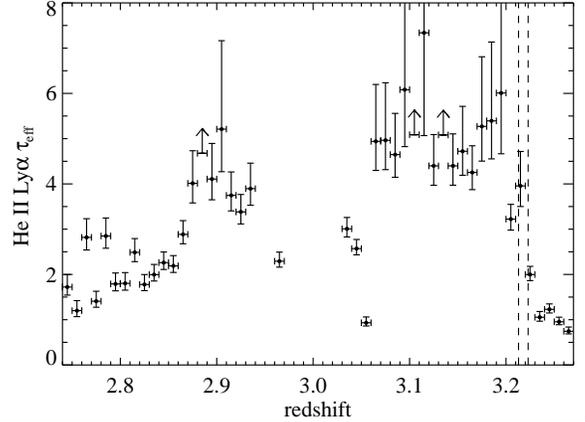}
\caption{$\tau^{\rm eff}_{\rm He\,II,\, \alpha}(z)$, calculated using COS G130M data in $\Delta z = 0.01$ bins. Regions contaminated by geocoronal \ion{N}{1}~$\lambda$1200 and \ion{H}{1}~Ly$\alpha$ are excluded. Confidence intervals are 68\% on the observed flux, but incorporate our full range of uncertainty in the continuum fit. Limits shown are from confidence interval calculations on the data, not from sensitivity calculations. The vertical dashed lines indicate possible lower edges for the line-of-sight proximity zone. The data shown are tabulated in Appendix~\ref{app:tab_data}.}
\label{fig:tau_z}
\end{figure}

Figure~\ref{fig:tau_z} shows the evolution of the \ion{He}{2}~Ly$\alpha$ with redshift, in $\Delta z = 0.01$ bins.
Such a bin size is large enough to show the general trend of evolution to higher optical depths at higher redshifts, while still showing important features such as the onset of the Ly$\alpha$ forest at $z \approx 2.87$, the transverse proximity effect at $z=3.05$, and the line-of-sight proximity effect at high redshift.
Regions with possible contamination by geocoronal emission have been removed.
The origin of lower-than-expected \ion{He}{2} optical depths at $z \approx 2.96$ and $3.03$ is unclear.
The former is also seen in STIS data \citep{heap00}, and seems real.
The latter is not seen in the STIS data, but lies outside the region that should experience any geocoronal contamination.
Given the flux asymmetry on the red and blue side of geocoronal Ly$\alpha$, it is unlikely to be due to contaminating flux in the wings of the COS LSF.

Our COS/G140L data confirm the optical depths derived from G130M, although with larger error bars due to the much shorter exposure time.
Unfortunately the $z \sim 3$ regions with interestingly high flux in G130M are contaminated by geocoronal emission in G140L.

The coverage probability quoted is for the source and background counts, assuming a Poissonian distribution and Feldman--Cousins confidence intervals \citep{feldman98}.
We use the full range of continuum parameters shown in Figure~\ref{fig:cos_full_continuum}, regardless of the confidence interval, because we cannot rigorously assign a coverage probability to the continuum range.

\subsection{Limits from the Low Density IGM}
\label{sec:low-density_tau}

The desire to distinguish between the absorption of the \ion{He}{2}~Ly$\alpha$ forest (largely matching the density fluctuations of the \ion{H}{1}~Ly$\alpha$ forest) and absorption from the most tenuous parts of the IGM (a Gunn-Peterson trough signalling incomplete reionization) has motivated previous studies comparing hydrogen and helium absorption.
Notably, \citet{zheng98a} used Q0302 itself to argue on the basis of the \citet{hogan97} FUV data that the absorption seen at higher redshifts was not simply a Ly$\alpha$ forest.

\begin{figure}
\epsscale{1.2}
\plotone{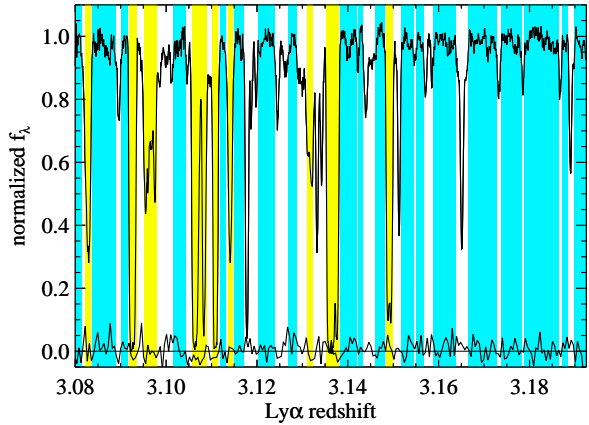}
\caption{Normalized VLT \ion{H}{1} spectrum of Q0302 (upper curve), showing low-density regions with very high transmission (light blue shading) and moderate- to high-density regions with $\tau_{\rm H\,I} > 0.4$ (yellow shading). The lower curve shows the normalized COS \ion{He}{2} data. All regions used in the analysis are shown in this figure.}
\label{fig:hi_regions}
\end{figure}

More recently, \citet{mcquinn09a} argued that looking at helium absorption in low-density regions can set surprisingly stringent limits on the \ion{He}{2} ionization factor, $x_{\rm He\,II}$.
There is a long, black trough in the spectrum of Q0302, from $z \approx 3.06$--3.19.
We look for high-transmission regions in the \ion{H}{1} spectrum to identify low density IGM, and calculate the $\tau^{\rm eff}_{\rm He\,II, \, \alpha}$ in  those regions.
We look for regions lying within noise and continuum-normalization uncertainty of 100\% transmission (approximately normalized $f_{\lambda} = F > 0.93$; shaded in light blue in Figure~\ref{fig:hi_regions}).
We require each region to be $\delta z \geq 0.001$, which corresponds observationally to four COS resolution elements and physically to one comoving Mpc at these redshifts.
We cover a total path length of $\Delta z = 0.0579$.

In the high-transmission (low-density) regions we find $\tau^{\rm eff}_{\rm He\,II, \, \alpha} = 4.84^{+0.89}_{-0.49}$ at 95\% confidence ($4.84^{+0.42}_{-0.29}$ at 68\% confidence), calculated as in the previous section.
Continuum uncertainty is a subdominant effect for here---the upper error bar changes only from $0.89$ to $0.81$ if no continuum uncertainty is included.
We follow \citet{mcquinn09a} in assuming that the baryon density (relative to the mean density) associated with the highest-transmission regions is $\Delta_b \sim 0.15$.
Using Equation~2 in \citet{syphers11b} for $\Delta_b = 0.15$, $\tau^{\rm eff}_{\rm He\,II, \, \alpha} = 4.84$ gives $x_{\rm He\, II} \approx 0.009$.
This is consistent with the roughly estimated lower limit of $\tau^{\rm eff}_{\rm He\,II, \, \alpha} \gtrsim 4$ \citet{mcquinn09a} derived using STIS data of Q0302, although the STIS data could not actually resolve the low-density regions.
The dominant error here is the assumption of the value of $\Delta_b$; the cosmological parameters used are a relatively negligible source of error.
By the scaling relation of \citet{mcquinn09a}, this result implies that the volume-averaged \ion{He}{2} fraction $x_{\rm He\,II, \, V}=0.04$ at mean density, although note that this changes to $0.025$ if $\Delta_b = 0.2$.
Due to systematic uncertainties in the normalization of high-resolution optical spectra, the value of $\Delta_b$ is difficult to determine exactly.
In addition, translating the low-density ionization fraction to the volume-averaged quantity relies on $\Gamma_{\rm He\,II}$ being constant on scales $\gtrsim$10~cMpc.
There are some reasons to believe this is true \citep{mcquinn09a}, but as shown in Section~\ref{sec:eta}, there may in fact be variation in the ionization rate on scales smaller than this.

These uncertainties do not affect the surprising detection of a finite optical depth in this fairly high-redshift region.
Some simulations and models have predicted potentially moderate optical depths here \citep{mcquinn09,dixon09}, but in practice these were not necessarily expected to be observable \citep{worseck11a}.
Effects from non-uniform ionizing backgrounds could be important here \citep[e.g.,][]{dixon13}.
Transverse proximity zones can produce dramatic flux spikes such as that seen near $z=3.05$ in Q0302, but low-luminosity and perhaps distant quasars could produce very small effects detectable only in a statistical sense.

This flux detection warrants further scrutiny because it is both interesting and possibly subject to nontrivial systematic errors.
For example, the background subtraction method used in this paper gives an estimate of 122 source counts in the redshift region in question, over all exposures.
The pipeline background subtraction method gives 190 source counts, and the difference gives some idea of the size of possible systematics.
However, we think our background subtraction method has substantially smaller systematics than the pipeline method.
Recall that the pipeline method, although somewhat improved recently (including CALCOS 2.18.5, used here) compared to earlier versions, still estimates the background from regions offset in the cross-dispersion direction rather than directly in the PSA.

We can make one check of the systematics by comparing the optical depth in the low-density regions with the optical depth we measure in regions of higher density.
Measuring a similar optical depth there would indicate that we are not seeing a signal from the IGM, and are instead seeing background modeling problems, scattered light, or other spurious signals.

The low-transmission (high-density) regions we use are shaded in yellow in Figure~\ref{fig:hi_regions}.
The cutoff we choose for ``high density'' is somewhat arbitrary.
The mean \ion{H}{1} optical depth $\tau_{\rm eff}(z \sim 3.1) \approx 0.4$ \citep{faucher-giguere08}.
Making a cut of $\tau > 0.4$ leaves us with enough data to perform our analysis, so we choose this.
The low-transmission ($\tau > 0.4$, $F < 0.67$) region has a total path length $\Delta z = 0.0260$.
As with the high-transmission region discussed above, we require a minimum of $\delta z \geq 0.001$ for each segment.

We detect no source counts in the high-density region; indeed, we find a very slightly unphysical result of about $-1$ source count with a background of about 542 counts.
Reassuringly, this is not unphysical enough to make us doubt our background model.
We therefore obtain $\tau^{\rm eff}_{\rm He\,II, \, \alpha} > 5.71$ (68\%) or $>5.00$ (95\%), including continuum uncertainty.
Our recommendation in \citet{syphers11b} was to quote two additional values in addition to the confidence interval when dealing with unphysical counts.
For the first we follow \citet{feldman98}, defining the sensitivity of an observation to be the average upper limit of observered counts for an experiment with the same background as ours, but with zero real signal.
Doing so, we find a sensitivity of $\tau_{\rm eff}=4.94$ at 95\% confidence ($5.49$ at 68\% confidence).
The second method is what we call the detector upper limit\footnote{\citet{kashyap10} refer to this somewhat opaquely as just the ``upper limit'', distinguished from the ``upper bound,'' or what we call the source upper limit. As in \citet{syphers11b}, we use frequentist confidence intervals for this calculation.}.
We follow \citet{kashyap10} in defining this as the maximum intensity a source can have without having a probability of $\beta$ of being detected at a significance level $1 - \alpha$.
Using a 95\% signficance level ($\alpha=0.05$) and $\beta = 0.5$, we get a detector upper limit of $\tau_{\rm eff}=5.15$ ($\tau_{\rm eff}=6.40$ at 68\% significance, $\alpha=0.32$.)
The sensitivity and detector upper limit demonstrate that we {\it should} have seen a signal like that in the high-transmission region ($\tau_{\rm eff} = 4.84$), if one were present in the low-transmission region.

This is a tantalizing result.
We have formally not only detected flux in the low-density region, but we have detected {\it no} flux in the high-density region, and the results in the two regions are different with moderate confidence.
(The precise confidence depends on method chosen, but is $>$90\% when one considers the range of possible $\tau_{\rm eff}$ values in the high-transmission region compared to sensitivities or detector upper limits.)
This is the first detection of flux in a \ion{He}{2} Gunn-Peterson trough apart from one with the ACS/SBC prism, averaging over many sightlines \citep{syphers11a}.
However, while the ACS/SBC prism background was extremely low, it was also very poorly understood.
In addition, the prism (resolution $R \sim 50$--300) could not exclude clear transverse proximity regions, let alone separately examine different density regions of the IGM.

Such an interesting result demands corroboration.
Unfortunately further observation of Q0302 with the current generation of instruments is unlikely to yield a more conclusive result.
While the COS G140L grating has reduced per-{\AA}ngstrom background compared to G130M, it does not have the resolution necessary to robustly distinguish high and low-density regions. 
More importantly, at this point our trust in the result is primarily limited by COS systematics, rather than having too few counts for statistical significance.
Verification with Ly$\beta$ is tempting, but the effective area of HST/COS is not sufficient for a reasonable exposure time in the Q0302 Ly$\beta$ trough ($\lambda \lesssim 1075$~\AA, excluding the proximity zone).

One important comparison that will need to be made is to SDSS0915, a \ion{He}{2} quasar with slightly higher redshift and flux than Q0302 \citep{syphers12}, which will probe the same redshifts at the same resolution.
{\it HST} observations of SDSS0915 were recently obtained (GO 12816, PI Syphers), and analysis is forthcoming.

\section{Helium-Ionizing Background}
\label{sec:eta}

\begin{figure*}
\epsscale{1.2}
\plotone{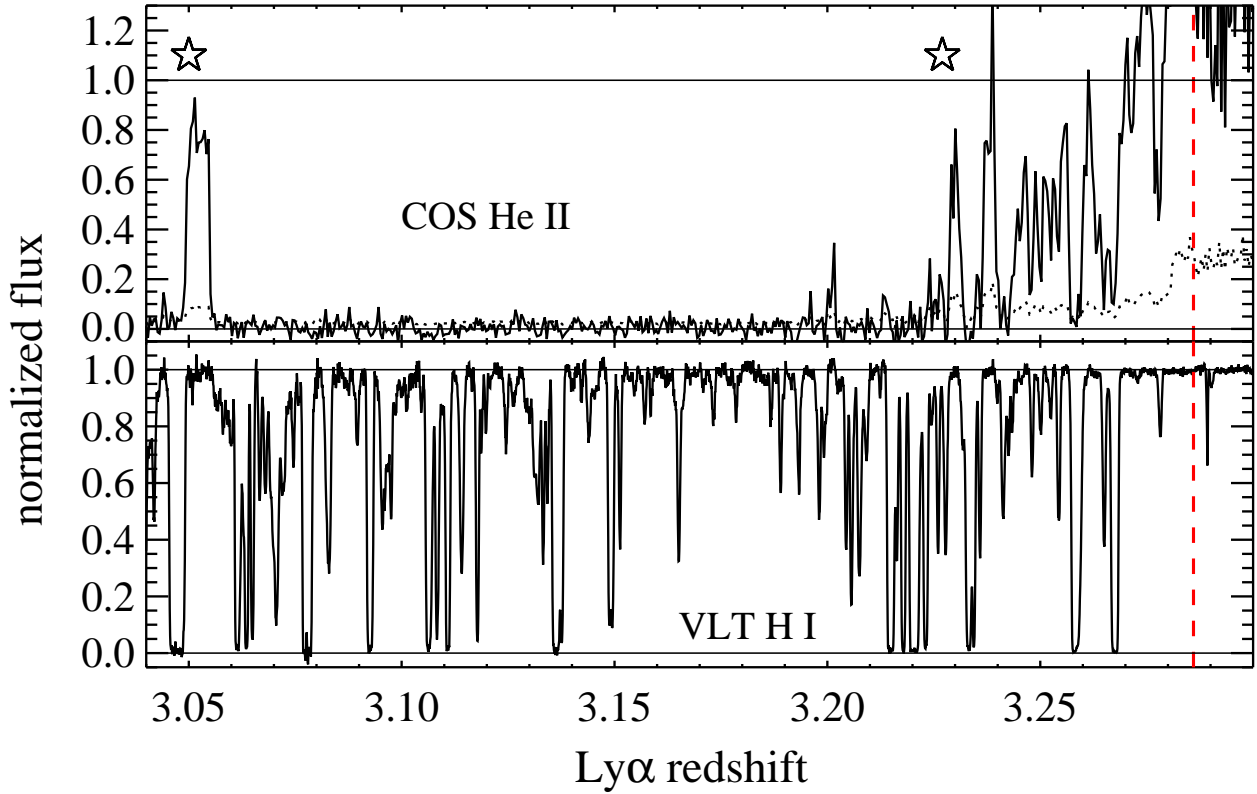}
\caption{Normalized COS/G130M data show the higher-redshift \ion{He}{2}~Ly$\alpha$ forest in the upper panel, while normalized VLT/UVES data show the \ion{H}{1}~Ly$\alpha$ forest in the lower panel. The VLT data are smoothed to two pixels (6~km~s$^{-1}$, about one resolution element), and the COS data are smoothed to 14 pixels (33~km~s$^{-1}$, about two resolution elements). Redshifts of known nearby quasars are indicated with stars, with redshift uncertainty comparable to the symbol width.}
\label{fig:cos_optical_comparison_highz}
\end{figure*}

\begin{figure*}
\epsscale{1.2}
\plotone{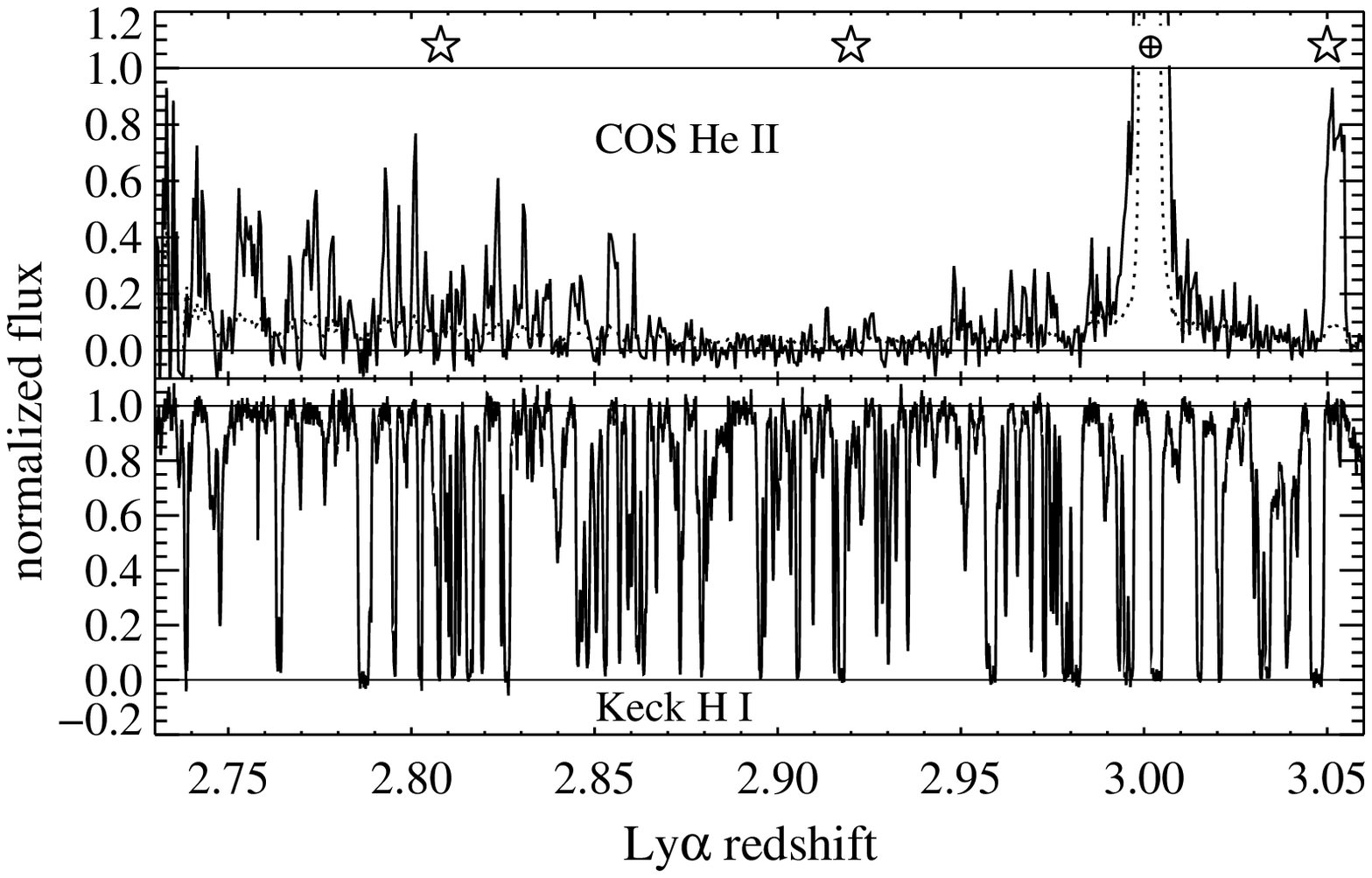}
\caption{Normalized COS/G130M data show the lower-redshift \ion{He}{2}~Ly$\alpha$ forest in the upper panel, while normalized Keck/HIRES data show the \ion{H}{1}~Ly$\alpha$ forest in the lower panel. The Keck data are smoothed to three pixels (8~km~s$^{-1}$, about one resolution element), and the COS data are smoothed to 14 pixels (36~km~s$^{-1}$, about two resolution elements). Geocoronal contamination due to \ion{H}{1}~Ly$\alpha$ is marked, with both wings shown to allow some estimate of its contamination at lower redshifts. The flux excess at $z \approx 2.97$ is also seen in the STIS data, and is probably real. Geocoronal \ion{N}{1} $\lambda$1200 and $\lambda$1134 emission has been removed by using night-only data. Redshifts of known nearby quasars are indicated with stars.}
\label{fig:cos_optical_comparison_lowz}
\end{figure*}

The hardness of the metagalactic UV background experienced by the IGM is of great interest, as this determines the ionization state of the photoionized IGM.
While variations in the \ion{H}{1}~Ly$\alpha$ forest alone are dominated by density fluctuations, combining this information with the \ion{He}{2}~Ly$\alpha$ forest breaks this degeneracy.
Metals can also be used, but the advantage of using \ion{He}{2} is that it is detectable at all points rather than only in rare complexes.
Figures~\ref{fig:cos_optical_comparison_highz} and \ref{fig:cos_optical_comparison_lowz} show both the \ion{H}{1} and \ion{He}{2} data.

The standard observational measure of hardness from comparing hydrogen to helium is defined as 

\begin{equation} \label{eqn:eta_def}
   \eta \equiv \frac{N_{\rm He\, II}}{N_{\rm H\, I}} \approx 4 \frac{\tau_{\rm He \, II}}{\tau_{\rm H \, I}}
\end{equation}

\noindent
where the factor of $4$ arises from the ratio of wavelengths, and holds if the lines are broadened purely by turbulence (including Hubble flow).

\subsection{Methods for Determining $\eta$}
\label{sec:eta_methods}

There is currently some debate about whether there are substantial $\eta$ fluctuations in the post-reionization epoch.
There are theoretical claims that such fluctuations are expected (\citealt{dixon13}; L.~Graziani et al.\ in prep.) and that they are not \citep{mcquinn13}.
Many observations see such $\eta$ variations by comparing the two forests using a variety of methods \citep[e.g.,][]{zheng04b,shull04,fechner07a}, although some of these methods may be noticeably biased \citep{mcquinn13}.
(We discuss this possible bias in detail in Appendix~\ref{app:eta}.)
Recently, significant $\eta$ variations have been seen in the post-reionization epoch by looking at metal-line systems \citep{agafonova13}.
This debate aside, there is no question that large $\eta$ fluctuations are expected prior to the completion of helium reionization, due to the patchy nature of the process and the rarity of the bright quasars responsible \citep[e.g.,][]{fardal98,furlanetto09a,compostella13}.

For calculating $\eta$, we use the forward-modeled estimator $\hat{\eta}_{\rm fm}$ \citep{heap00,fechner07a,mcquinn13}.
In Appendix~\ref{app:eta} we discuss this method in detail, including its own biases, and compare it to the method using the ratio of effective optical depths.
Here we summarize our implementation.
To find $\hat{\eta}_{\rm fm}$, we take the normalized optical \ion{H}{1} spectrum and smooth it to about a single resolution element (two pixels for VLT, three for Keck).
This spectrum is converted into a line-resolved $\tau_{\rm H \, I}(z)$, which is then converted to an estimated $\hat{\tau}_{\rm He \, II}(z)$ and from there an estimated transmission $\hat{T}_{\rm He \, II}(z)$.
This estimated helium spectrum is convolved with the COS LSF, and then compared to the actual helium flux measurements in specified redshift bins.
We choose $\hat{\eta}_{\rm fm}(z)$ as those values giving the closest flux matches to the real helium data.
We use the normalized VLT spectrum where possible, at $z_{\alpha} > 2.956$, and Keck data at lower redshifts, down to $z_{\alpha} \sim 2.73$, where our COS coverage ends.

For calculating the helium fluxes, we sum the COS signal counts, $n_s$, and the background counts, $n_b$, measured in each exposure, completely excluding regions contaminated by geocoronal emission.
We calculate $\tau_{\rm He \, II}$ using Feldman-Cousins confidence intervals \citep{feldman98}.
For lower limits on $\tau_{\rm He \,II}$ when $n_s < 0$, we use the sensitivity as defined above.

The coverage of the error bars is nominally 68\%, but in fact it is not well defined for all the data.
We use 68\% confidence intervals for the \ion{He}{2} source counts, but we use the full range of possible continua.
The latter error has ill-defined coverage, but is likely much higher than 68\%.
Fortunately it is a subdominant error.
Errors on $\hat{\eta}_{\rm fm}$ are taken comparing the hydrogen data to the upper and lower limits of the helium flux in each bin.

We make the standard assumption of pure turbulent line broadening.
There is some evidence that this is a good assumption \citep{zheng04b}, although other analysis claims thermal broadening may be important over substantial regions of the forest \citep{fechner07a}.
This question cannot be answered with highly absorbed \ion{He}{2} data, and low-redshift data comparing the hydrogen and helium Ly$\alpha$ forests is the best approach.
The analyses of such data thus far have been limited to {\it FUSE} observations, where poor S/N and background subtraction make the comparison difficult.
Our upcoming COS/G130M observations of the brightest two \ion{He}{2} quasars, HE2347$-$4342 and HS1700$+$6416, should allow resolved study of the forest at good S/N (GO 13301, PI Shull).
This will be ideal for ascertaining the importance of thermal broadening.

Because of the much lower average optical depth in \ion{H}{1}, continuum normalization errors become the dominant source of error in the optical spectra.
We are averaging over relatively large bins for this high-resolution data, and the S/N is good, so the random errors are quite small.
However, binning does not help the systematic errors, which might reach a few percent in the normalization of high-resolution data at $z \sim 3$ \citep{faucher-giguere08}.
(See Appendix~\ref{app:continuum} for further discussion of this possible effect, as well as why we do not trust simulations to accurately determine the offset.)
As an illustration of how a systematically low continuum might change our results, we determine $\hat{\eta}_{\rm fm}$ using both the standard normalized continuum and a continuum that lies 2\% higher \citep[approximately the $z \sim 3$ continuum systematic estimate in][]{faucher-giguere08}.
Since this just changes $\tau_{\rm H \, I}$ to $\tau_{\rm H \, I} + 0.02$ to a good approximation, there is a noticeable effect only for regions with fairly low \ion{H}{1} optical depth, but this is a large portion of the $z \sim 3$ IGM.

Contaminating absorption is unlikely to be an issue.
In the absence of Lyman-limit systems, intervening Galactic or lower-redshift IGM absorbers affect very little of the spectrum.
For example, SDSS shows a substantial overdensity of galaxies at $z \sim 0.03$ near the Q0302 sightline.
There are 19 spectroscopically verified galaxies clustered in $0.0276 < z < 0.0316$ within 30$'$ ($\sim$1~Mpc) of the Q0302 sightline, with a mean redshift $z = 0.02916 \pm 0.00028$ and a group dispersion of 340~km~s$^{-1}$.
It is likely that we see a \ion{Si}{2}~$\lambda$1304.37 absorption line from this group, in which case \ion{Si}{2}~$\lambda$1260.42 absorption could contaminate the proximity profile.
However, other absorption is dominant in any reasonably large redshift bin, and in that particular region of the spectrum, continuum normalization (of both hydrogen and helium) causes the dominant errors.

\subsection{$\eta$ Results}
\label{sec:eta_results}

\begin{figure}
\epsscale{1.2}
\plotone{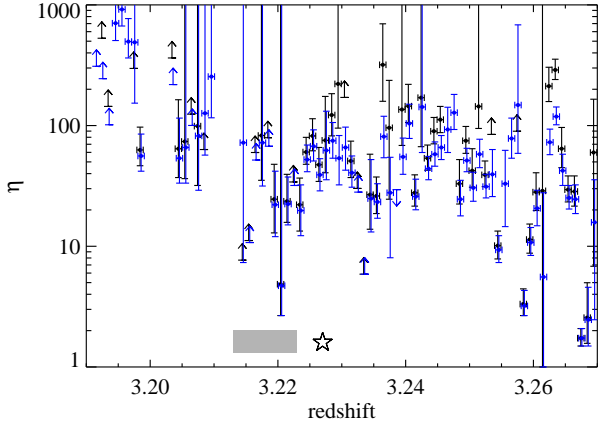}
\caption{High-redshift $\hat{\eta}$ in $\Delta z = 0.001$ bins. Both sets of points show the forward-modeled $\hat{\eta}_{\rm fm}$ (see text for details). Black points show the estimator for the standard normalization of the \ion{H}{1} spectrum while the blue points show what happens if we increase the continuum by 2\%. The errors on $\eta$ are 68\% confidence intervals. Points where $\hat{\eta}$ cannot be determined due to continuum-normalization problems are not plotted. A slight offset in $z$ between black and blue points has been introduced in the plot for clarity. The gray box shows the range of possible lower edges for the line-of-sight proximity zone, and the star marks the redshift of a known quasar (discussed in Section~\ref{sec:trans_prox}).}
\label{fig:eta_fm_highz}
\end{figure}

\begin{figure}
\epsscale{1.2}
\plotone{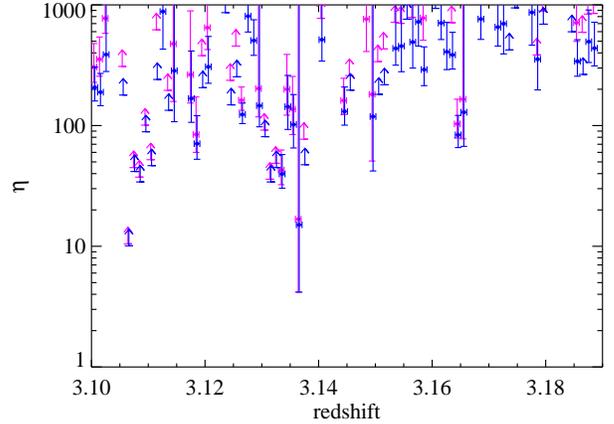}
\caption{Intermediate/high-redshift $\hat{\eta}$ in $\Delta z = 0.001$ bins, covering the Gunn-Peterson trough at $z=3.10$--3.19. The methodology and notation is the same as Figure~\ref{fig:eta_fm_highz}, except that black points are not shown because the default \ion{H}{1} continuum is problematic over some of this redshift range. Magenta points show estimates where normalized flux values greater than one have been replaced with values slightly less than one (see text for further discussion).}
\label{fig:eta_fm_gp}
\end{figure}

\begin{figure}
\epsscale{1.2}
\plotone{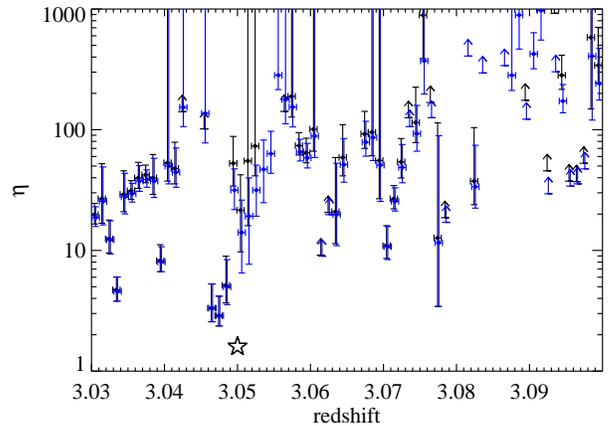}
\caption{Intermediate-redshift $\hat{\eta}$ in $\Delta z = 0.001$ bins, covering the $z=3.05$ transverse proximity quasar (redshift indicated with a star). The methodology and notation is the same as Figure~\ref{fig:eta_fm_highz}. Note the transverse proximity zone covers $z \approx 3.050$--3.055, as discussed in Section~\ref{sec:trans_prox}. The region of very low $\eta$ just below $z=3.05$ is associated with a strong \ion{H}{1} absorption line.}
\label{fig:eta_fm_midz}
\end{figure}

In the era before full helium reionization, including in the proximity zone, we see strongly varying $\eta$ (Figure~\ref{fig:eta_fm_highz}).
The region at $z>3.27$ has no usable data, possibly because of intrinsic quasar line emission (Section~\ref{sec:intrinsic_emission}).
In the Gunn-Peterson trough, $\eta$ varies, but is generally $\eta \gtrsim 100$, sometimes much higher (Figures~\ref{fig:eta_fm_gp} and \ref{fig:eta_fm_midz} as well).
The transverse proximity zone at $z=3.05$ is associated with moderately lower $\eta \sim 50$, although with considerable uncertainty.
There is a region at $z=3.046$--3.049, just below the transverse proximity zone, that is quite hard, with $\eta < 10$, although there are possible effects from systematics when $\tau_{\rm H\, I}$ is very large.
This \ion{H}{1} absorption is associated with a metal complex, which \citet{heap00} model and find could be exposed to a low-luminosity AGN background, although not a background as hard as suggested by the $\eta$ values.
Our two optical data sets overlap for the redshift range in Figure~\ref{fig:eta_fm_midz}, and we reassuringly find that they yield $\eta$ estimates that agree, despite their independent normalization.

For comparison to \citet{heap00}, we note that ``softness ratio,'' defined by as the ratio of photoionization rates of \ion{H}{1} and \ion{He}{2}, is related to the \ion{He}{2}-to-\ion{H}{1} abundance ratio by the formula \citep{fardal98}

\begin{equation}
   S \equiv  \frac {\Gamma_{\rm H\, I}} {\Gamma_{\rm He\, II}} \approx \frac{n_{\rm H\,II}}{n_{\rm He\,III}} \frac{\alpha^{(A)}_{\rm H\,I}}{\alpha^{(A)}_{\rm He\,II}} \frac{N_{\rm He\,II}}{N_{\rm H\,I}} \approx 2.26 \, \left( \frac{T}{\rm 15,000\, K} \right)^{-0.047}  \,  \eta
\end{equation}

\noindent
where the first approximation assumes photoionization equilibrium, and the second approximation assumes high ionization (so $n_{\rm H\,II}/n_{\rm He\,III} \approx n_{\rm H}/n_{\rm He}$, the elemental abundance ratio).
The dependence on temperature is derived for $T$ ranging from 8,000--20,000~K.
Since this assumes high ionization, it applies in the proximity zones and the post-reionization era, but at the redshifts in question it should also be an acceptable approximation in the Gunn-Peterson trough, if our ionization values of Section~\ref{sec:low-density_tau} are correct.

\begin{figure}
\epsscale{1.2}
\plotone{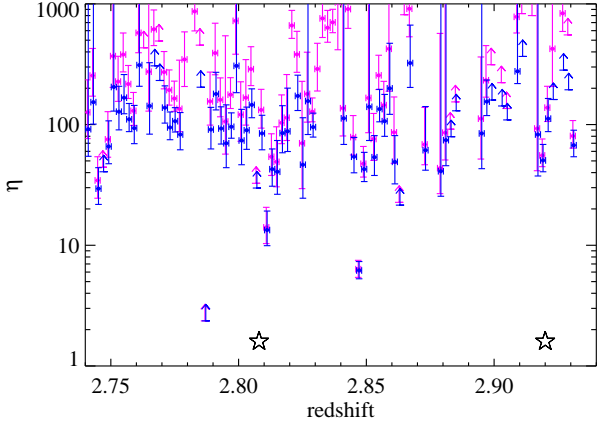}
\caption{Low-redshift $\hat{\eta}$ in $\Delta z = 0.002$ bins, including the post-reionization recovery to a Ly$\alpha$ forest (compare to Figure~\ref{fig:cos_optical_comparison_lowz}). The methodology and notation is the same as Figure~\ref{fig:eta_fm_gp}. The redshifts of known quasars are marked with stars.}
\label{fig:eta_fm_lowz}
\end{figure}

\citet{heap00} find a soft radiation field for the metagalactic background at higher redshifts ($S \sim 800$--1000, or $\eta \sim 350$--450), and a somewhat harder field at the $z \sim 3.05$ transverse proximity zone and lower redshifts ($S \sim 120$, or $\eta \sim 50$).
Our results provide broad agreement with these results (Figures~\ref{fig:eta_fm_highz}--\ref{fig:eta_fm_midz}), although we prefer a softer spectrum at the lowest redshifts (Figure~\ref{fig:eta_fm_lowz}).

In the LOS proximity zone, \citet{heap00} estimate that $\eta$ steadily decreases with increasing redshift, averaging $\eta \sim 200$ in the proximity zone and reaching $\eta \sim 30$--40 very near the quasar.
We find lower $\eta$ inside the proximity zone, as expected, but we do not find the steady decrease when approaching the quasar (Figure~\ref{fig:eta_fm_highz}).
\citet{hogan97} find an even harder $\eta \sim 20$ plausible in much of the LOS proximity zone, and a significantly higher value ($\eta \gtrsim 500$) in the Gunn-Peterson trough.
However, they were unable to determine these values well, due in part to a difficult and imperfect background subtraction of the GHRS data.

In the low-redshift regime (Figure~\ref{fig:eta_fm_lowz}), it is difficult to determine $\eta$ due to normalization uncertainty in the \ion{H}{1} data.
Many points have normalized fluxes $F>1$, which prohibits determination of moderate-to-large $\eta$ values (discussed further in Appendix~\ref{app:eta}).
As a result, most points do not have well-defined $\hat{\eta}$ for the default continuum, and so we do not plot this in Figure~\ref{fig:eta_fm_lowz}.
Instead we plot the points with 2\% continuum offset, as before, and additionally plot points where we have replaced all points with $F>1$ with $F=0.995$ (without shifting the continuum overall).
The choice of the flux replacement value is arbitrary, beyond the requirements that it must be less than one but very close to one.
Both of these methods introduce a bias toward lower $\eta$ compared to the default continuum, but in different ways.
The difference between blue and magenta points gives some idea of the systematic uncertainty, including dependence on the precise value of the replacement flux.
Figure~\ref{fig:eta_fm_gp} does not have continuum normalization problems at the same level, but we follow the same plotting there to maximize the number of useable points in this figure.
It is sparsely populated because there are many regions in the Gunn-Peterson trough where no $\eta$ or even reliable limit on $\eta$ can be found ($\eta > 1000$ estimates are not shown, as they are not reliable given the systematics).

Because we cannot determine $\eta$ at every redshift, we cannot present a robust determination of the average $\eta$ or the distribution of $\eta$ as some earlier works have attempted \citep[e.g.,][]{shull10}.
It is likely that in many regions without an $\eta$ estimate one can use 1000 as a lower limit, but because of the limitations of the forward-modeling method discussed in Appendix~\ref{app:eta}, this is not guaranteed.

\section{Proximity Effects}
\label{sec:proximity}

\subsection{Line-of-Sight Proximity Effect}
\label{sec:los_prox}

Quasars ionize large regions of the surrounding IGM, affecting the average transmission of Ly$\alpha$ photons, as seen in Figure~\ref{fig:cos_optical_comparison_highz}.
In this section we explore the line-of-sight (LOS) proximity effect of Q0302 itself, and in the next section we consider transverse proximity effects of lower-redshift quasars.

The size of an ionized region can be found by solving the following equation adapted from \citet{cen00},

\begin{equation} \label{eqn:he_ionization_evolution}
\frac{dR^3_i}{dt} = 3 H(z) R^3_i + \frac{3 \dot{N}_{\gamma}}{4 \pi \langle n_{\rm He \, II} \rangle \Delta_b} - C_{\rm He\,III} \langle n_{e} \rangle \Delta_b \alpha_R R^3_i
\end{equation}

\noindent
where $\dot{N}_{\gamma}$ is the emission rate of \ion{He}{2}-ionizing photons, the electron and helium density averages are cosmic mean, $\Delta_b = \rho_b / \langle \rho_b \rangle$ is the baryon overdensity, $C_{\rm He\,III}$ is the clumping factor, and $\alpha_R$ is the recombination coefficient (we use case A).
Here $\Delta_b$ refers to large-scale average overdensity, and is thus not simply related to $C$.
Near quasars $\Delta_b > 1$, but for larger ionization zones $\Delta_b \approx 1$, because cosmic mean density is quickly reached as one moves away from a quasar.
Because we are working with helium rather than hydrogen, we can no longer assume $n_{\rm ion} = n_e$ as \citet{cen00} did, and thus modify their equation to distinguish them.
It is reasonable to take \ion{H}{1} and \ion{He}{1} as fully ionized by this epoch, so $n_e = n_{\rm H} + n_{\rm He} + n_{\rm He} (1 - x_{\rm He\, II})$.

The solution to Equation~\ref{eqn:he_ionization_evolution} is

\begin{eqnarray}
R_i(t) &=& \bigg\{ \frac{3 \dot{N_{\gamma}}}{4 \pi \langle n_{\textrm{He}} \rangle x_{\rm He\,II} \Delta_b \cdot (3 H(z) - C_{\rm He\,III} \langle n_{e} \rangle \Delta_b \alpha_R)} \cdot \notag \\
& & \big[ \exp \left( [3 H(z) - C_{\rm He\,III} \langle n_{e} \rangle \Delta_b \alpha_R)] \cdot t \right) - 1 \big] \bigg\}^{1/3} \label{eqn:he_ionization_solution}
\end{eqnarray}

\noindent
For simplicity, we have taken $H(z)$ and $C$ as constants, which is a good approximation for realistic quasar lifetimes of tens of Myr.

There are two limiting cases of interest.
First, where recombinations are happening very quickly ($C n_e \Delta_b \alpha_R \gg H(z)$), in which case the classic Str\"{o}mgren sphere solution balancing ionization with recombination is recovered.
This equilibrium is not reached for quasars at $z \sim 3$ \citep[e.g.,][]{donahue87}.
The second is when recombinations are negligible and the quasar lifetime is short compared to the Hubble time ($t \ll 1/H(z)$), so expansion can be neglected.
In this case the number of photons produced is equal to the number of \ion{He}{2} ions in the sphere, and the solution used in \citet{cen00} is recovered:

\begin{eqnarray}
R_i &=& \left( \frac{3 \dot{N}_{\gamma} t}{4 \pi \langle n_{\rm He} \rangle x_{\rm He\,II} \Delta_b} \right)^{1/3} \notag \\
 &\approx& 13.7 \left( \frac{\dot{N}_{\gamma}}{10^{57} \; {\rm s}^{-1}} \frac{t}{10 \; {\rm Myr}} \frac{1}{x_{\rm He\,II}} \right)^{1/3} \left( \frac{4}{1+z} \right) \label{eqn:he_ionization_radius_simple}
\end{eqnarray}

\noindent
where in the approximation we have scaled to WMAP9 parameters and set $\Delta_b = 1$.
Distances are in proper Mpc.
If $x_{\rm He\,II} \ll 1$, this implies a nontrivial metagalactic ionizing background, which changes the appropriate definition of the edge of the ionization zone.

Q0302 has a substantial proximity zone, where there is detectable \ion{He}{2}~Ly$\alpha$ transmission, and the helium and hydrogen lines are strongly correlated (Figure~\ref{fig:cos_optical_comparison_highz}).
It may extend down to $z \approx 3.223$ or even $z \approx 3.213$, which would be $13.5$ (or $15.8$) proper Mpc.
Also interesting is the small but likely real transmission spike at $z=3.201$.

This suggests a long luminous phase for Q0302, to have ionized such a large region.
Assuming a power-law flux with the EUV spectral index $\alpha_{\nu} = -0.82$ found in Section~\ref{sec:specfit}, we estimate a rate of helium-ionizing photons of about $1.4 \times 10^{57}$~s$^{-1}$, although this could range from 1.2 to 3.6$\, \times 10^{57}$~s$^{-1}$ considering the uncertainty in $\alpha_{\nu}$.
Here we use an observed flux at the \ion{He}{2} edge of $7.4 \times 10^{-16}$~erg~s$^{-1}$~cm$^{-2}$~\AA$^{-1}$, where we have removed IGM pLLS absorption.
Variability can also contribute to the uncertainty; variability observed in the optical suggests flux changes by $\sim$10\% (Section \ref{sec:trans_prox}).
FUV variability has not been observed, but is certainly possible (Section \ref{sec:specfit}).

Because several relevant parameters have large uncertainty, we present two extreme cases.
For a lower limit on $t$, we take a radius of $13.5$~Mpc, $\dot{N_{\gamma}} = 3.6 \times 10^{57}$~s$^{-1}$, $C=3$, $T = 2 \times 10^4$~K, $\Delta_b=1$, and $x_{\rm He\,II} = 0.05$.
The quasar only needs to be on for $0.17$~Myr in this scenario to send out sufficiently many ionizing photons (recombinations are negligible).
For an upper limit, we take a radius of $15.8$~Mpc, $\dot{N_{\gamma}} = 1.2 \times 10^{57}$~s$^{-1}$, $C=10$, $T = 10^4$~K, $\Delta_b=1.5$, and $x_{\rm He\,II} = 1$.
In this case, the quasar must be luminous for 31~Myr to ionize the region, and recombinations are important.
A middle ground of the most plausible parameters suggests the luminous phase of Q0302 has lasted a few million years, typical for a bright quasar.

To more precisely model the quasar age based on the ionization zone, numerical models with realistic density fields, heating and cooling, wavelength-dependent cross sections, a nonzero metagalactic UV background, and other effects should be used (see W.~Zheng et al.\ 2014, in prep., for an example).
But the uncertainty in the input parameters shows why this is not a worthwhile exercise for Q0302, particularly because its LOS proximity zone is likely complicated by a transverse proximity effect of a $z=3.23$ quasar, discussed in the next section.

While many \ion{He}{2} quasars show proximity zones of various sizes \citep{syphers12}, the large LOS proximity zone of Q0302 is contrasted with what is seen in the two brightest \ion{He}{2} quasars.
HS1700$+$6416 has a small zone \citep{syphers13}, and HE2347$-$4342 has no proximity zone at all, likely due both to a poorly understood infalling absorber and also to having only recently entered its luminous phase \citep{fechner04,shull10}.
Newer targets are yielding interesting information from their LOS proximity zones (W.~Zheng et al.~in prep.), showing the advantage of moving beyond the two best-studied \ion{He}{2} quasars to a statistical sample.

In addition to the transmission of \ion{He}{2}, we also see the proximity zone in $\eta$.
Inside the proximity zone, $\eta$ is smaller than in the Gunn-Peterson trough (Figure~\ref{fig:eta_fm_highz}), although we do not see strong evidence for any trend in $\eta$ across the ionization zone.

\subsection{Transverse Proximity Effects}
\label{sec:trans_prox}

The ionization zones around lower-redshift quasars can intersect the line of sight, creating a transverse proximity effect.
The presence or absence of such zones can reveal information about the lifetimes and beaming angles of quasars \citep[e.g.,][]{adelberger04,furlanetto11}.
With information on both \ion{H}{1} and \ion{He}{2} absorption, we can break the degeneracy with density that has hampered attempts to see a transverse proximity effect in the \ion{H}{1} forest alone.
We can also get information on the hardness of the SED responsible for the ionization.

The Dobrzycki--Bechtold void is a long region of very low \ion{H}{1} optical depth at $z \approx 3.17$--3.185 \citep{dobrzycki91}.
The FUV spectrum revealed that the void could not be due to a nearby quasar, as \ion{He}{2} has a high optical depth in this region (Figure~\ref{fig:cos_optical_comparison_highz}) and the ionizing spectrum is soft (Figure~\ref{fig:eta_fm_gp}).

Contrasted with this is the transmission region we find at $z_{\rm center}=3.0521$, which was seen (though not resolved) in the STIS data \citep{heap00}.
\citet{jakobsen03} associated this with the transverse proximity zone of a $z=3.050 \pm 0.003$ quasar, so we find no evidence for a redshift offset, and indeed an even closer coincidence in redshift than originally thought.
Isotropic emission from the quasar is therefore fully consistent with the data.
In projection, this quasar is $6 \farcm 49$ from Q0302, or $3.05$ transverse proper Mpc, and has SDSS magnitudes $g=20.68$, $r=20.31$.
In this region we find $\tau^{\rm eff}_{\rm He\,II, \, \alpha} = 0.30^{+0.16}_{-0.10}$ (95\%), with an optical depth uncertainty dominated by continuum uncertainty.
This region has a very flat plateau in flux, and choosing just that plateau we find $\tau^{\rm eff}_{\rm He\,II, \, \alpha} = 0.27^{+0.16}_{-0.10}$ (95\%).
The redshift range for this plateau is $z=3.0496$--$3.0545$, which is about 4.7 comoving Mpc.
Using STIS data, \citet{heap00} overestimated this length by more than a factor of three, and using this measurement would lead one overestimate the lifetime of the quasar.
With the correct transverse proximity zone size, we estimate a quasar lifetime of $>$15.5 Myr.
However, the uncertainty in the quasar redshift is large enough to give about a 3~Myr uncertainty on the limit.
We confirm that the weak \ion{C}{4} absorption system at $z=3.04725$ is extremely well aligned with a strong \ion{H}{1} and \ion{He}{2} absorber, and not the void seen in both \ion{H}{1} and \ion{He}{2}, and thus may not accurately measure the foreground quasar radiation.

We also see a previously undiscovered flux signature that may be from the transverse proximity effect of a much brighter quasar, although the detection of the effect is not conclusive.
Q0301$-$005 is a luminous quasar (SDSS $g=17.95$, $r=17.65$) that is $22 \farcm 92$ from Q0302, or 10.6 transverse proper Mpc \citep{barbieri86}.
There are three SDSS spectra of this target, spanning about 90 days in the quasar frame.
We adopt a redshift of $z=3.227 \pm 0.001$ based on measuring the low-ionization lines \ion{O}{1}/\ion{Si}{2}~$\lambda$1305 and \ion{C}{2}~$\lambda$1335 in all three SDSS spectra, and using the [\ion{O}{3}]-corrected rest wavelengths of \citet{vanden-berk01}.
The systematic uncertainty in the redshift is somewhat larger.
Using the lab rest wavelengths for the low-ionization lines, one obtains $z=3.232 \pm 0.002$.
The redshift from [\ion{O}{3}]-corrected wavelengths is slightly preferred by other lines, including Ly$\alpha$, but the redshift from lab wavelengths aligns slightly better with the transmission feature.

There is a clear excess in transmission at this redshift in the helium data at $z = 3.228$--$3.232$ (Figure~\ref{fig:cos_optical_comparison_highz}), above what might be expected the LOS proximity effect alone, although it is difficult to disentangle increased ionizing flux from decreased density.
The amount of the spectrum affected by this transverse proximity zone is unclear, and complicated by line absorption.
However, nonzero flux in the helium spectrum is visible down to $z=3.201$, so it is possible that this transverse proximity zone is quite large, and it likely overlaps the LOS proximity zone of Q0302 at higher redshift.

Because the helium transmission at this redshift rises above that of some very low-density regions at higher redshifts closer to Q0302, we find it likely that Q0301$-$005 is responsible for much of the flux in this region.
However, Q0302 itself should have a nontrivial contribution, and its ionizing radiation precludes a definitive assignment of any features to Q0301$-$005.
From the SDSS spectra we find that Q0301$-$005 has a continuum luminosity $L_{1450} = \lambda L_{\lambda}(1450 {\rm \AA}) \approx (1.56 \pm 0.10) \times 10^{47}$~erg~s$^{-1}$, where the uncertainty reflects the observed variability between spectra.
(We use $E(B-V)=0.102$ for Q0301$-$005; \citealt{schlafly11}.)
Q0302 is somewhat more luminous, with $L_{1450} \approx (1.98 \pm 0.22) \times 10^{47}$~erg~s$^{-1}$, but it is also slightly farther away (12.5 proper Mpc vs.\ 10.6 for Q0301$-$005).
Without knowing the EUV spectral index of Q0301$-$005 we cannot precisely determine the helium-ionizing flux ratios, but the two quasars will contribute approximately equally at $z \sim 3.23$. 

The existence of a transverse proximity zone aligned in redshift with this quasar at such a distance would imply that it has been quite luminous for $>$34~Myr.
Because of the good alignment between the quasar redshift and the transmission redshift, the lifetime is likely longer than this lower limit.
This limit is somewhat higher than those seen in other direct quasar lifetime tests using proximity effects in \ion{He}{2}, $\eta$, metal lines, or fluorescent Ly$\alpha$ emitters.
All these lifetime tests yield lower limits, and the largest such lower limit previously seen is 25~Myr, with many much lower \citep{jakobsen03,worseck07,goncalves08,trainor13}.

As Q0301$-$005 is bright and has been known for a long time, suggestions that it {\it could} cause a transverse proximity effect are not new.
However, there has never before been credible evidence for the effect.
\citet{worseck06} present three pieces of evidence when making this suggestion, but none hold up under scrutiny.
First, they use an $\hat{\eta}_{\rm bin}$ estimator to find a single low $\eta$ point near the quasar, but this estimator is flawed \citep[Appendix~\ref{app:eta};][]{mcquinn13} and nearby points do not agree.
Second, they use a variation of forward modeling, but they do not actually cover the redshift of Q0301$-$005.
Third, they suggest that the apparent LOS proximity zone of Q0302 is too large to be caused by Q0302 alone, and likely needs Q0301$-$005 to enlarge it.
While this combination of ionization zones likely {\it is} happening, Q0302 alone is entirely sufficient to create a zone with the measured radius, even assuming very short lifetimes (Section~\ref{sec:los_prox}).
The STIS data of \citet{worseck06} were insufficient to show the \ion{He}{2} transmission feature discussed above.

Other transverse proximity effects are possible.
At lower redshifts, Figure~\ref{fig:eta_fm_lowz} shows two points where $\eta$ is unusually low, $z=2.811$ with $\hat{\eta}_{\rm fm} = 14^{+7}_{-4}$ and $z=2.847$ with $\hat{\eta}_{\rm fm} = 6.2^{+1.3}_{-1.0}$.
These points have hydrogen and helium fluxes that are neither saturated nor near the continuum, and thus should be well measured with believable confidence intervals.
As mentioned in Section~\ref{sec:fuv_data}, the wavelength alignment should be within a single resolution element, and each $\Delta z = 0.002$ bin contains about 10 resolution elements, so there should be no systematics from wavelength calibration.
These $\eta$ features therefore may be real, although they would be more convincing if broader.
There is a known quasar at $z=2.808$ \citep{worseck06}, at a transverse separation $d = 11 \farcm 23$ (5.40 Mpc) and magnitude $r = 21.3$.
No known quasars lie near the higher-redshift low-$\eta$ point.

One additional nearby quasar is known, Q0302-003~D113 at redshift 2.92 \citep{steidel03}.
This quasar is quite faint, with $R_{\rm AB}=24.64$, and thus assuming an average quasar spectrum of $\alpha_{\nu}=-0.5$ above 1000~\AA\ and $\alpha_{\nu}=-1.5$ below, a helium-ionizing photon rate $\dot{N_{\gamma}} \simeq 8 \times 10^{53}$~s$^{-1}$.
Despite its faintness, it may still have an effect because it lies only $4\farcm88$ from the Q0302 sightline, or about 2.33~Mpc.
A quasar of this luminosity could have an ionization zone reach this distance in about 60~Myr if $x_{\rm He\,II} \sim 1$, and much more quickly if helium is largely ionized already.
There may be some slightly increased \ion{He}{2} transmission at this redshift (Figure~\ref{fig:cos_optical_comparison_lowz}), and there may be a locally lower $\eta$ (Figure~\ref{fig:eta_fm_lowz}).
If this detection is real, it offers only a weak constraint on quasar lifetime because of the close proximity to the Q0302 sightline, and the large uncertainty of ages derived by means other than causality.

\citet{worseck06} performed a slitless spectroscopic survey of the Q0302 sightline, so we have very complete coverage out to $\sim$10$'$ and a limiting magnitude of $V \approx 22$.
Such surveys are very helpful where possible, but the most interesting transverse proximity effects are those from bright quasars fairly far from the sightline \citep{furlanetto11}.
Although such quasars are rare, our detection of the \citet{barbieri86} quasar is proof that this can be done, and yields interesting limits.

\section{Conclusion}
\label{sec:conclusion}

We have presented the COS/G130M \ion{He}{2} spectrum of Q0302$-$003 at resolving power $R \sim 18$,000, the first $z>3$ quasar observed in the FUV at $R > 3000$.
This allows us to resolve many interesting features, from accurately determining the nature of the $z \approx 3.05$ transverse proximity effect to resolving lines in the LOS proximity zone.
It also allowed us to find a possible new transverse proximity effect due to a quasar that is luminous and relatively distant from the sightline.
We infer a luminous phase of $>$34~Myr if this feature is caused by the transverse proximity quasar.

The broad evolution of \ion{He}{2} effective optical depth with redshift matches expectations and what is seen in other sightlines, with a \ion{He}{2}~Ly$\alpha$ forest of $\tau_{\rm eff} \sim 2$ by $z \sim 2.8$ turning into a Gunn-Peterson trough of $\tau_{\rm eff} \gtrsim 4$ by $z \sim 2.9$.
Long black troughs indicate incomplete reionization at higher redshifts.

Looking more closely, the COS/G130M resolution allowed us to contrast the effective optical depths of \ion{He}{2} in low- and high-density regions of the IGM.
We detect flux in the low-density regions and no flux in the high-density regions, indicating a \ion{He}{2} fraction of about 1\% in the low-density IGM at $z \sim 3.1$--3.2, and perhaps a few percent in the volume-averaged IGM.
Although the detection is statistically significant at $>$95\%, and the difference between it and the non-detection is significant at $>$90\%, confirmations from other sightlines would allow more confidence that systematics are ruled out.

Comparing hydrogen to helium absorption allows us to trace fluctuations in the ionizing background, parametrized by $\eta = N_{\rm He\, II} / N_{\rm H\, I}$.
In Q0302's LOS proximity zone we find somewhat lower $\eta$ values than previous work, indicating a harder spectrum, and compatible with the $\alpha_{\nu}=-0.82$ intrinsic EUV spectral index we fit.
Prior to the end of helium reionization, we see strong $\eta$ fluctuations, as expected.
The post-reionization regime is difficult to constrain due to normalization issues with the \ion{H}{1} data, but it seems noticeably harder (lower $\eta$) than the pre-reionization Gunn-Peterson trough, as some models predict.

These observations of Q0302 show what is possible with higher-resolution \ion{He}{2} data.
Recent and upcoming COS/G130M observations of five additional \ion{He}{2} quasars, including three not observed in detail before and one at higher redshift, will allow us to draw conclusions both more detailed and more statistically valid than has been heretofore possible.

\acknowledgments

We thank Tae-Sun Kim for providing the continuum-normalized VLT spectrum, and George Becker for providing the continuum-normalized Keck spectrum.
JMS thanks the Institute of Astronomy, Cambridge University, for their support through the Sackler visitor program.
This work was supported by NASA grants NNX08AC146 and NAS5-98043 and the Astrophysical Theory Program (NNX07-AG77G from NASA) at the University of Colorado at Boulder.

{\it Facilities:} \facility{HST (COS)}, \facility{ARC (TripleSpec)}

\appendix

\section{Estimating $\eta$}
\label{app:eta}

Calculating the column density of \ion{He}{2} compared to that of \ion{H}{1} can reveal basic information on the metagalactic UV background (UVB) due to the different ionization potentials, breaking the degeneracy that absorption strength has between the UVB and local density.
The best method for doing this has recently been the subject of debate, and so we investigate the question in this appendix.

Perhaps the most obvious method to calculate $\eta$, considering its definition in terms of column densities (Equation~\ref{eqn:eta_def}), is to perform line fitting on both the hydrogen and helium spectra.
Of course, this is only possible in the post-reionization epoch, where it makes sense to think of absorption lines rather than a Gunn-Peterson trough.
Such fitting been attempted along the HE2347$-$4342 sightline \citep{zheng04b}, but the poor quality of the {\it FUSE} data used made the conclusions uncertain.
Upcoming COS G130M/1222 data (GO 13301) should allow us to do this in a limited redshift range post-reionization.

There are two other estimators one can use for $\eta$.
The first is inspired by the approximate equivalency of $\eta$ to the ratio of optical depths in Equation~\ref{eqn:eta_def}.
We define $\hat{\eta}_{\rm bin} = 4 \tau^{\rm eff}_{\rm He \, II} / \tau^{\rm eff}_{\rm H \, I}$ with the effective optical depths calculated over some redshift bin size (this is $\hat{\eta}_{\rm simple}$ in \citealt{mcquinn13}).
This method has the advantage of being easily defined whether or not the absorption can be sensibly thought of as lines.
Its primary disadvantage is that, owing to resolution and S/N issues, we must approximate optical depths with effective optical depths, which can bias the results.
Obviously this problem can be much worse in COS/G140L data (resolution $\sim$140~km~s$^{-1}$) than in COS/G130M (resolution $\sim$17~km~s$^{-1}$), but it also depends on the redshift bin size chosen.

The second estimator, $\hat{\eta}_{\rm fm}$, uses ``forward modeling,'' starting with the \ion{H}{1} spectrum and using some $\eta$ to turn this into a predicted \ion{He}{2} spectrum, as described here in Section~\ref{sec:eta}.
This is the method of Section~3.1 of \citet{heap00}, Eqn.~1 in \citet{fechner07a}, and Eqn.~2 of \citet{mcquinn13}.
(We note in passing that to avoid censoring data when the logarithms are taken, the operations should be performed using complex numbers, to allow for optical depths of negative flux values.)
The resulting predicted spectrum is compared to the actual transmission in the helium spectrum, and $\hat{\eta}_{\rm fm}$ is chosen in each redshift bin to produce the best match.
In the limit that the hydrogen and helium spectra approach constants in a given bin, $\hat{\eta}_{\rm bin} \to \hat{\eta}_{\rm fm}$ and they both give the correct value.

When accounting for the LSF of the UV data (the LSF of the optical data is negligible by comparison), with $\hat{\eta}_{\rm bin}$ one must convolve prior to binning and finding $\tau_{\rm H \, I}$.
With $\hat{\eta}_{\rm fm}$ one can convolve prior to modeling a predicted helium spectrum \citep[as in][]{fechner07a} or after modeling \citep[as in][]{heap00,mcquinn13}.
The latter preserves the line-resolved optical depths of the \ion{H}{1} data, and is therefore the better choice.
The COS LSF has large non-Gaussian wings \citep{kriss11}, so the effect of convolution is larger than might be naively expected.

Due to using effective optical depths, $\hat{\eta}_{\rm bin}$ is a biased estimator, tending to underestimate $\eta$.
To illustrate this effect we take the cosmological simulation of \citet{smith11}, with a 50$h^{-1}$~cMpc box and $1024^3$ cells, so an intrinsic resolution of about $R$$\approx$~55,000 ($5.45$~km~s$^{-1}$) at $z \sim 3$.
This is sufficient to have line-resolved optical depths, and therefore calculate an accurate $\eta$ value\footnote{This simulation matches the basic properties of the $z \sim 3$ IGM Ly$\alpha$ forest well, but we caution that it was not intended to model helium reionization because a homogenous ionizing background was adopted.
We only rely on it to illustrate how a binned $\eta$ estimator is biased, using a realistic cosmological density field, and are not focused on the true values or distributions of $\eta$.
}.
We take a slice of the simulation at $z=3$, and send 500 rays through the box to create simulated spectra.

\begin{figure}
\epsscale{0.7}
\plotone{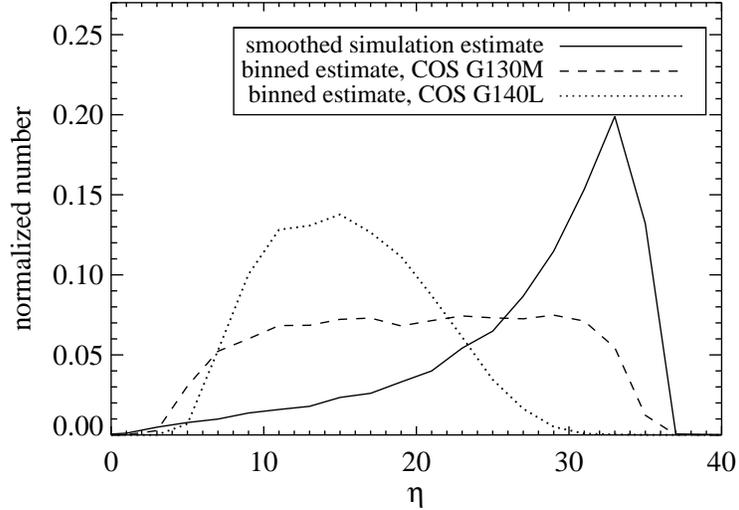}
\caption{The binned $\eta$ estimator in $\Delta z=0.001$ ($75$~km~s$^{-1}$) bins. The solid line shows the ``true'' $\eta$ distribution from the simulation, smoothed to $8.6$~km~s$^{-1}$ to average over unphysically narrow features (see text). The dashed line shows the measured $\hat{\eta}_{\rm bin}$ distribution as measured after the \ion{He}{2} spectrum has been convolved with the COS/G130M LSF. The dotted line shows the same for COS/G140L. Note that the simulation distribution of $\eta$ is not intended to represent the true $\eta$ distribution at $z \sim 3$. This plot serves only to illustrate the qualitative changes introduced in the measured distribution by resolution effects.}
\label{fig:eta_bin_bias}
\end{figure}

Figure~\ref{fig:eta_bin_bias} shows how resolution can affect the measurement of $\eta$, leading to values biased low.
This figure is illustrative rather than a quantitative analysis of the effect, as it neglects two very important effects that would reduce the fidelity to the true $\eta$ distribution---no noise is included (a nontrivial effect for \ion{He}{2} spectra) and no continuum-normalization uncertainty is included (a large systematic for \ion{H}{1} spectra).
Note that the shape of the ``true'' $\eta$ distribution used does not matter for the present discussion, which compares the ideal behavior of the estimators on noise-free data.
If one attempts to simulate the real $\eta$ distribution and compare these quantitatively to the observations, noise must be taken into account.

In this simulation $\eta(z)$ varies rapidly, down to the smallest scales resolved ($\Delta z \approx 7 \times 10^{-5}$), which we believe is not real---this is smaller than the thermal broadening scale, for example, and \citet{mcquinn13} claim most of the $\eta$ variation occurs on scales larger than 2 cMpc.
Substantial variation on much smaller scales dramatically reduces the fidelity of $\hat{\eta}_{\rm bin}$ compared to what it would have in real application, so we smooth over the unphysically rapid variations by convolving the data with a narrow Gaussian (${\rm FWHM} = 8.6$~km~s$^{-1}$, half the G130M FWHM, and sufficiently small to keep real features resolved).
The degree of fidelity also depends on the bin size chosen, when the bin size is larger than the instrumental resolution.
Here we have used $\Delta z = 0.001$ as we did on the real data in Section~\ref{sec:eta_results}; larger bins will reduce fidelity in this example, where rapid oscillations are important.
However, fidelity is good for all bin sizes used in this paper (including $\Delta z = 0.002$), as discussed below.

\begin{figure}
\epsscale{0.7}
\plotone{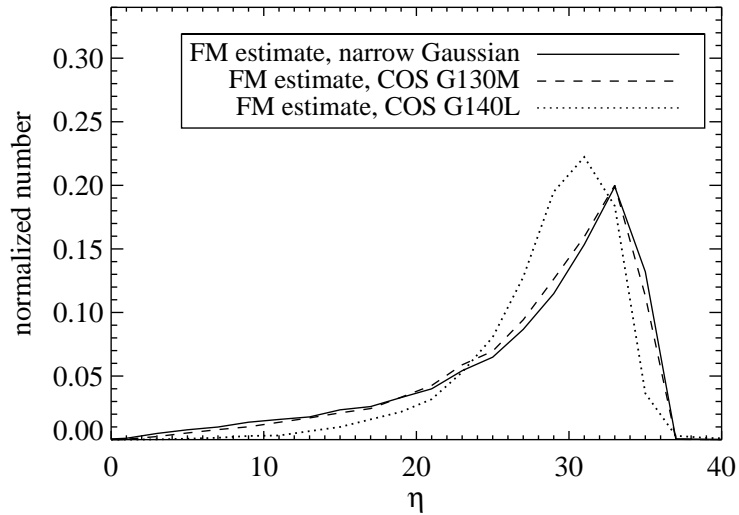}
\caption{The forward-modeled $\eta$ estimator in $\Delta z=0.001$ ($75$~km~s$^{-1}$) bins. The solid line is the ``true'' $\eta$ distribution from the simulation, smoothed to $8.6$~km~s$^{-1}$ to average over unphysically narrow features (see text). The dashed line shows the measured $\hat{\eta}_{\rm fm}$ distribution as measured on simulated COS/G130M data, while the dotted line shows the same on COS/G140L data. No noise is included in the simulations.}
\label{fig:eta_fm_bias}
\end{figure}

\begin{figure}
\epsscale{0.7}
\plotone{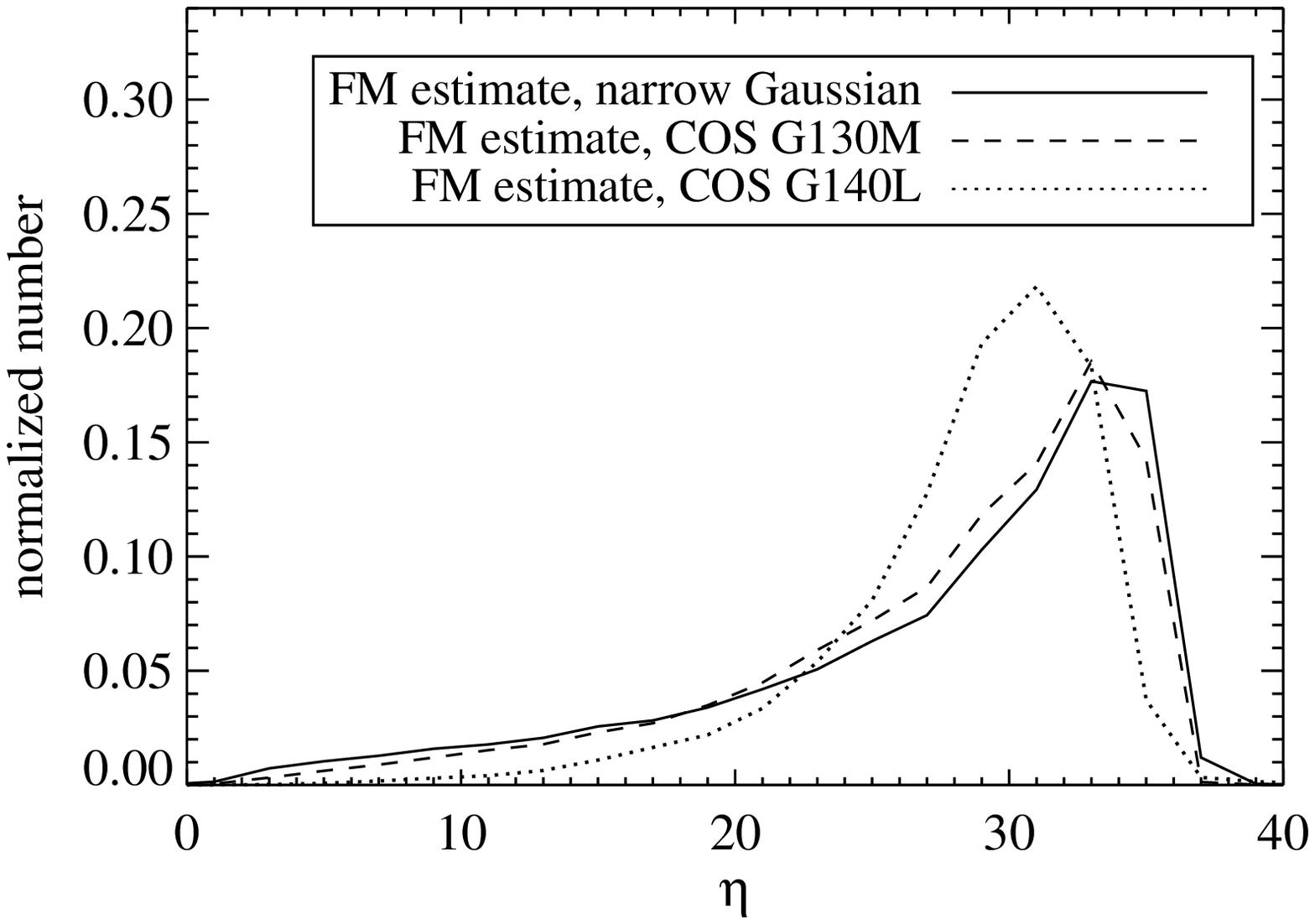}
\caption{The same as Figure~\ref{fig:eta_fm_bias}, except using $\Delta z=0.0005$ ($37$~km~s$^{-1}$) bins. The nearly identical results compared to the previous figure indicate that we are choosing bin sizes small enough to be converged. A similar test shows this to be true for bins of $\Delta z=0.002$ as well.}
\label{fig:eta_fm_bias_0005}
\end{figure}

The performance of $\hat{\eta}_{\rm fm}$ on the same simulation is shown in Figures~\ref{fig:eta_fm_bias} and \ref{fig:eta_fm_bias_0005}.
Here we show simulated COS/G130M and COS/G140L data, and also data convolved with the same narrow Gaussian.
The effect of the non-Gaussian wings of the COS LSF can sometimes be substantial in certain bins, but this figure shows that it has little effect on the overall distribution.
This shows that $\hat{\eta}_{\rm fm}$ on G130M data maintains high fidelity to a realistic version of the simulation, while on G140L data it is slightly biased.
Comparison of Figures~\ref{fig:eta_fm_bias} and \ref{fig:eta_fm_bias_0005}, which use bins of $\Delta z = 0.001$ and $\Delta z = 0.0005$ respectively, shows that nearly identical results are obtained.
A similar test for $\Delta z = 0.002$ shows that all the bin sizes used in this paper are small enough to converge (although tests indicate that $\Delta z = 0.005$ bins are {\it not}).

In real application $\hat{\eta}_{\rm fm}$ is also biased in many bins, toward high values.
By definition, $\hat{T}_{\rm He \, II}(\eta)$ should be monotonically decreasing as $\eta$ increases, and it should be straightfoward to find the $\eta$ that results in this being equal to the measured $T_{\rm He \, II}$.
However, it is important to note that this will not be the case anywhere near the hydrogen continuum ($T_{\rm H \, I} \approx 1$), where there will be occasional noise spikes above unity (which we emphasize will occur even in perfectly well-behaved and well-normalized data).
For small $\eta$, $\hat{T}_{\rm He \, II}$ will behave as expected, but when $\eta$ approaches a critical value, the points where $T_{\rm H \, I} < 1$ will saturate at $\hat{T}_{\rm He \, II}=0$.
A noise spike above unity in $T_{\rm H \, I}$ will increase $\hat{T}_{\rm He \, II}$ without bound as $\eta$ increases, however, eventually coming to dominate no matter how small its measure in wavelength.
This can lead to two different $\eta$ values that minimize $| \hat{T}_{\rm He \, II}(\eta) - T_{\rm He \, II} |$, and even the real crossing at the smaller $\eta$ will be pulled higher, so that $\eta_{\rm true} < \hat{\eta}_{\rm fm}$.
In cases where the true value of $\eta$ is large, no modeled $\eta$ will acceptably minimize the difference.
The simple solution of ignoring all spikes above a normalized flux of one will bias results, however, even in cases where the $\eta$ estimate is well below the deviation from monotonicity.
These are the $F > 1$ problems discussed in Section~\ref{sec:eta_results}, and is one reason we encourage comparison of $\eta$ determined both with and without a continuum offset, to assess the impact of this bias.

\section{Unresolved absorption in the IGM}
\label{app:continuum}

Another effect addressed in \citet{mcquinn13} is the precise determination of the true optical continuum for calculating \ion{H}{1} optical depths.
This can have a considerable effect in some redshift bins even for very high S/N optical data, because of systematic continuum offsets due to the low-density IGM \citep{faucher-giguere08}.
\citet{mcquinn13} implement a method to find the continuum of the quasar HE2347$-$4342, which they describe as an improvement over standard normalization techniques.
However, they do caution that it does not have general applicability---indeed, due to more structure in the continuum, even the very high S/N observation of HS1700$+$6416 (one of the brightest quasars on the sky at $z \sim 3$) is not amenable to this method.

In addition, there are concerns about how well we can actually determine this systematic offset using simulations.
If present, this offset in the continuum placement most likely arises from a large number of weak, unresolved \ion{H}{1} absorbers, whose frequency per unit redshift, $d{\cal N}/dz$, is sufficiently high to produce what appears as continuous opacity.
One can estimate this effect using the effective optical depth model, as modeled by \citet{fardal98}.  For the Ly$\alpha$ forest of absorbers at $z \approx 3$, ranging in \ion{H}{1} column density from $N_1 < N < N_2$,

\begin{equation}
  \frac {d \tau_{\rm eff}} {dz} = \int_{N1}^{N2} \frac {d^2 {\cal N}} {dz \, dN} \, 
      [1 - \exp(-\tau(N)) ]  \, dN   \;  . 
\end{equation}  

To estimate $\tau_{\rm eff}$, we adopt the bivariate distribution in column density and redshift, $(d^2 {\cal N} / dz \, dN) = (A/N_r)(N/N_r)^{-\beta} (1+z)^{\gamma}$, where $N_r = 10^{17}$~cm$^{-2}$ is a reference column density at which $\tau \approx 1$.
The parameters $(A, \beta, \gamma)$ are chosen to be (0.0669, 1.49, 2.58) from models A2 and A3 of \citet{fardal98}.
Using the weak-line limit, $[1 - \exp(-\tau)] \approx \tau$, with Gunn-Peterson line optical depth, $\tau = (\pi e^2 / m_e c)(N f \lambda / c)$, the integral can be performed analytically,

\begin{equation}
  \frac {d \tau_{\rm eff}} {dz} = \frac {A N_r}{(2-\beta)} (1+z)^{\gamma} 
       \left( \frac {\pi e^2}{m_e c} \right) \left( \frac {f \lambda}{c} \right) 
       \left[ \left( \frac {N_2}{N_r} \right) ^{2-\beta} - 
         \left(  \frac {N_1}{N_r} \right) ^{2-\beta} \right]    \;  ,
\end{equation}  

\noindent
and applied to hydrogen Ly$\alpha$ ($f = 0.4162$, $\lambda = 1.216 \times10^{-5}$~cm) at redshift $z \approx 3$ to give the relation

\begin{equation}
  \frac {d \tau_{\rm eff}} {dz} \approx (2.10) 
       \left[ \left( \frac {N_2}{N_r} \right) ^{0.51} - 
       \left( \frac {N_1}{N_r} \right)^{0.51} \right]    \;  . 
\end{equation}  

In one instead uses model A1 ($A=0.145$, $\beta=1.40$, $\gamma=2.58$) at $z=3$, the coefficient is $3.87$, although $\tau_{\rm eff}$ changes only by 18\%.
More recent surveys of the Ly$\alpha$ forest \citet{kim02,rudie13} generally agree, but with factors of two variations in this result.
For example, in their extensive survey at $\langle z \rangle = 2.4$ ($12.0 < \log N < 17.0$), \citet{rudie13} find $\beta = 1.65 \pm 0.02$ and a normalization at $10^{-13}$~cm$^{-2}$ that is a factor of 2.5 lower than models A2 and A3.
Uncertainty in the redshift dependence, $\gamma$, may affect the result by a similar factor \citep{kim02}.

We see that weak absorbers, in the range $12.0 < \log N < 13.0$, could contribute an amount $\Delta (\tau_{\rm eff}) \approx 0.005$--$0.03$ to the effective line opacity at $z \approx 3$.  
However, the precise amount of this extra opacity remains uncertain owing to the extrapolation of the $N^{-\beta}$ distribution to \ion{H}{1} column densities below those currently observable ($\log N < 12.3$), and the degree to which the flattening or turnover at $\log N \lesssim 13$ is real.
Any contribution of absorbers at $\log N < 12.0$ is particularly sensitive to a change in $\beta$, although such absorbers do not dominate.
Even with the extreme assumption of no evolution in $\beta$, including absorbers with $10.0 < \log N < 12.0$ changes $\tau_{\rm eff}$ by less than a factor of two, and this is an upper estimate.

Simulations in this column density regime are either poorly constrained by the data, or at the lower end, unconstrained.
In this Appendix we demonstrate that the uncertainty arising from measured and estimated forest contributions is considerable, and we do not think numerical simulations can produce trustworthy results to the desired accuracy in low-density IGM.
We conclude the placement of the true continuum must be regarded as a systematic uncertainty.

\section{Tabulated Data}
\label{app:tab_data}

Table \ref{tab:tau_heii_data} shows the data presented in Figure~\ref{fig:tau_z}.
Confidence intervals and lower limits are 68\% on the observed flux, but incorporate our full range of uncertainty in the continuum fit.

\begin{deluxetable}{cc}
\tablecolumns{9}
\tablewidth{0pc}
\tabletypesize{\footnotesize}
\tablecaption{Q0302 \ion{He}{2} Effective Optical Depth in $\Delta z = 0.01$ Bins}
\tablehead{
\colhead{Central redshift} & \colhead{$\tau^{\rm eff}_{\rm He \, II}$} \\
\colhead{} & \colhead{}}
\startdata
2.745 & $1.72_{-0.18}^{+0.28}$ \\
2.755 & $1.20_{-0.13}^{+0.22}$ \\
2.765 & $2.82_{-0.28}^{+0.41}$ \\
2.775 & $1.41_{-0.13}^{+0.22}$ \\
2.785 & $2.85_{-0.27}^{+0.40}$ \\
2.795 & $1.79_{-0.15}^{+0.24}$ \\
2.805 & $1.80_{-0.15}^{+0.24}$ \\
2.815 & $2.49_{-0.21}^{+0.30}$ \\
2.825 & $1.78_{-0.14}^{+0.22}$ \\
2.835 & $2.00_{-0.14}^{+0.22}$ \\
2.845 & $2.26_{-0.15}^{+0.24}$ \\
2.855 & $2.19_{-0.14}^{+0.23}$ \\
2.865 & $2.89_{-0.21}^{+0.30}$ \\
2.875 & $4.01_{-0.43}^{+0.72}$ \\
2.885 & $>$4.68 \\
2.895 & $4.11_{-0.46}^{+0.79}$ \\
2.905 & $5.21_{-0.94}^{+1.95}$ \\
2.915 & $3.75_{-0.34}^{+0.52}$ \\
2.925 & $3.38_{-0.27}^{+0.39}$ \\
2.935 & $3.90_{-0.37}^{+0.56}$ \\
2.945 & \nodata \\
2.955 & \nodata \\
2.965 & $2.29_{-0.13}^{+0.20}$ \\
2.975 & \nodata \\
2.985 & \nodata \\
2.995 & \nodata \\
3.005 & \nodata \\
3.015 & \nodata \\
3.025 & \nodata \\
3.035 & $3.01_{-0.18}^{+0.25}$ \\
3.045 & $2.57_{-0.14}^{+0.20}$ \\
3.055\tablenotemark{a} & $0.93_{-0.07}^{+0.13}$ \\
3.065 & $4.94_{-0.64}^{+1.26}$ \\
3.075 & $4.96_{-0.65}^{+1.27}$ \\
3.085 & $4.65_{-0.50}^{+0.91}$ \\
3.095 & $6.08_{-1.26}^{+\infty}$ \\
3.105 & $>$5.09 \\
3.115 & $7.34_{-2.27}^{+\infty}$ \\
3.125 & $4.40_{-0.43}^{+0.69}$ \\
3.135 & $>$5.08 \\
3.145 & $4.40_{-0.43}^{+0.71}$ \\
3.155 & $4.72_{-0.53}^{+0.99}$ \\
3.165 & $4.25_{-0.38}^{+0.59}$ \\
3.175 & $5.27_{-0.77}^{+1.54}$ \\
3.185 & $5.39_{-0.84}^{+1.74}$ \\
3.195 & $6.01_{-1.34}^{+\infty}$ \\
3.205 & $3.22_{-0.24}^{+0.33}$ \\
3.215 & $3.96_{-0.46}^{+0.76}$ \\
3.225\tablenotemark{a} & $2.00_{-0.13}^{+0.18}$ \\
3.235\tablenotemark{a} & $1.05_{-0.09}^{+0.13}$ \\
3.245\tablenotemark{a} & $1.23_{-0.08}^{+0.12}$ \\
3.255\tablenotemark{a} & $0.95_{-0.06}^{+0.10}$ \\
3.265\tablenotemark{a} & $0.74_{-0.06}^{+0.09}$
\enddata
\tablenotetext{a}{Clearly affected by either line-of-sight or transverse proximity effects.}
\label{tab:tau_heii_data}
\end{deluxetable}

\end{document}